\documentclass[aps,prx,twocolumn,showpacs,superscriptaddress,amsmath,amssymb]{revtex4-2}
\usepackage{graphicx}
\usepackage{latexsym}
\usepackage{bm} 
\usepackage{color}
\usepackage{epsfig}
\usepackage{multirow}
\usepackage{xcolor}
\usepackage{colortbl}
\usepackage{hhline}
\usepackage{simplewick}
\usepackage{soul}

\usepackage{graphicx}
\usepackage[colorlinks=true,linkcolor=blue,citecolor=blue,urlcolor=blue]{hyperref}
\usepackage{bbold}
\usepackage{gensymb}

\AtBeginDocument{%
    \newwrite\bibnotes
    \def\bibnotesext{Notes.bib}
    \immediate\openout\bibnotes=\jobname\bibnotesext
    \immediate\write\bibnotes{@CONTROL{REVTEX42Control}}
    \immediate\write\bibnotes{@CONTROL{%
    apsrev42Control,author="08",editor="1",pages="0",title="0",year="1"}}
     \if@filesw
     \immediate\write\@auxout{\string\citation{apsrev42Control}}%
    \fi
}%

\def \mbf {\mathbf}

\newcommand{\bra}[1]{\langle#1|}
\newcommand{\ket}[1]{|#1\rangle}
\newcommand{\braket}[3]{\langle#1|#2|#3\rangle}
\newcommand{\innp}[2]{\langle#1|#2\rangle}

\newcommand{\epvl}[1]{\langle#1\rangle}
\newcommand{\Epvl}[1]{\left\langle#1\right\rangle}
\def \Tr {\mathrm{Tr}}
\newcommand{\comment}[1]{}

\definecolor{mygreen}{rgb}{0, 0.7, 0}

\begin{document}

\title{Ultrafast optical control of charge orders in kagome metals}

\author{Yu-Ping Lin}
\affiliation{Department of Physics, University of California, Berkeley, California 94720, USA}
\author{Vidya Madhavan}
\affiliation{Department of Physics and Materials Research Laboratory, University of Illinois Urbana-Champaign, Urbana, Illinois 61801, USA}
\author{Joel E. Moore}
\affiliation{Department of Physics, University of California, Berkeley, California 94720, USA}
\affiliation{Materials Sciences Division, Lawrence Berkeley National Laboratory, Berkeley, California 94720, USA}

\date{\today}

\begin{abstract}
We show that ultrafast optical pump pulses provide effective control over charge orders in the kagome metals $A$V$_3$Sb$_5$ with $A=$ K, Rb, and Cs. Starting from the real charge density waves (rCDWs) at the $p$-type Van Hove singularity, we conduct a thorough analysis of the post-pump dynamics by time-dependent Hartree-Fock theory. Our analysis uncovers distinct dynamical phenomena under linearly and circularly polarized pumps. Linearly polarized pumps induce directional preferences in the rCDWs, accompanied by an enhancement in the flat band. Unexpectedly, charge nematicity also emerges and receives maximal enhancement at a resonant pump frequency, which we understand with a Rabi-oscillation-like model. On the other hand, circularly polarized pumps suppress the rCDWs uniformly and triggers imaginary CDWs (iCDWs) with charge loop currents. Our results can be directly compared to the pump-probe experiments on the kagome metals $A$V$_3$Sb$_5$.
\end{abstract}

\maketitle

\addtocontents{toc}{\string\tocdepth@munge}

\textit{Introduction.---}Recent advances in material synthesis have uncovered a large family of unconventional kagome materials \cite{yin22n,wang23nrp}. The vanadium-based kagome metals $A$V$_3$Sb$_5$ with $A=$ K, Rb, and Cs have received particular attention due to their abundant correlated phenomena \cite{neupert22np,jiang22nsr,wilson24nrm}. Real charge density waves (rCDWs) develop below $80$-$105$ K, which primarily exhibits intralayer $2\times2$ tri-hexagonal (TrH) charge bond orders (in CsV$_3$Sb$_5$, star-of-David order may coexist in some layers) \cite{kiesel13prl,wang13prb,tan21prl,denner21prl,lin21prb,park21prb,christensen21prb,dong23prb,tazai23nc,profe24ax,fu24ax}. Meanwhile, imaginary CDWs (iCDWs), which manifest in charge loop currents (CLCs), are proposed as possible time-reversal symmetry breaking states \cite{denner21prl,lin21prb,park21prb,christensen22prb,dong23prb,fu24ax}. Additional lattice symmetry breaking and even superconducting transitions occur at low temperatures. Despite these observations of a wide range of unconventional phenomena, the mechanisms behind them remain elusive. One of the major questions asks whether the rCDWs are driven by Coulomb repulsion or electron-phonon coupling, both of which are important in these materials.

Ultrafast optical techniques have emerged as powerful tools for the resolution of mechanism-related problems in correlated materials \cite{orenstein12pt,giannetti16aip,delatorre21rmp,boschini24rmp}. In particular, pump-probe experiments observe the relaxation dynamics after a short-pulse excitation, and coherent oscillations can reveal the relevant driving forces. Various pump-probe approaches have been applied to the kagome metals $A$V$_3$Sb$_5$ \cite{grandi24ax}, including coherent phonon spectroscopy \cite{ratcliff21prm,wang21prb,yu23prb}, time-resolved optical polarization-rotation measurement \cite{wu22prb}, time- and angle-resolved photoemission spectroscopy (trARPES) \cite{azoury23pnas,zhong24ax}, and time-resolved X-ray diffraction measurements (trXRD) \cite{ning24ax}. These experiments generally find THz coherent oscillations of phonon-related phase modes, indicating strong electron-phonon coupling in the rCDWs. However, whether the phonon-related amplitude modes exist remains inconclusive, as the frequency softening of THz modes has only been observed in one trARPES \cite{azoury23pnas}. The frequency range of the amplitude modes is a crucial indicator of whether the rCDW is driven by Coulomb repulsion or electron-phonon coupling. On the other hand, optical techniques can also be effective in controlling the correlated phases, especially when the competing orders are abundant as in the kagome metals $A$V$_3$Sb$_5$. In a recent laser-coupled STM experiment on RbV$_3$Sb$_5$ \cite{xing24n}, the promise of controlling the directional preference of the rCDWs is confirmed using linearly polarized pump pulses.

\begin{figure}[t]
\centering
\includegraphics[scale=1]{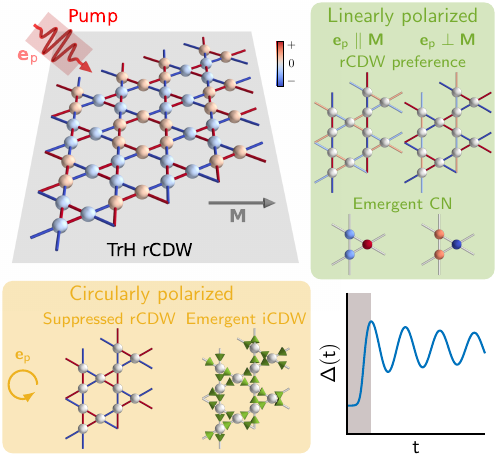}
\caption{\label{fig:feature} Schematic of controlling kagome metals with ultrafast optical pumps. (Upper left) The represented ground state of the kagome metals $A$V$_3$Sb$_5$ is a TrH rCDW. Under a pump pulse, the state undergoes nonequilibrium dynamics. (Lower right) After the short pulse duration (gray time domain), the charge orders $\Delta(t)$ evolve with strong collective-mode oscillations and possible shifts in the time averages. Choosing a linearly or circularly polarized pump, with the polarization vector $\mbf e_\text{p}$ pointing along a fixed or rotating direction, controls the relevant charge orders in the post-pump dynamics. In the charge orders, the site and bond colors indicate the density deviations from average (see the color bar), while the bond arrows represent the currents.}
\end{figure}

In this work, we explore the great potential of ultrafast optical techniques in understanding and controlling the kagome metals $A$V$_3$Sb$_5$ (Fig.~\ref{fig:feature}). The accessibility of multiple competing orders  makes these materials a logical starting point for the development and testing of theoretical approaches of broader applicability. We analyze the post-pump dynamics of the Coulomb-repulsion-driven charge orders at the filling $n_\text{f}=5/12$ on the kagome lattice. Since the kagome metals $A$V$_3$Sb$_5$ are nonmagnetic, we adopt a spinless-fermion Hubbard model to avoid complication from potential spin orders. Our analysis adopts the time-dependent Hartree-Fock theory, which is sufficiently versatile for a thorough investigation of driven dynamics. Remarkably, we find distinct dynamical phenomena under linearly and circularly polarized pumps. The linearly polarized pumps induce directional preferences of the rCDW, accompanied by an enhancement in the flat band. The charge nematicity (CN), where sublattice density imbalance breaks the rotation symmetry, also emerges and receives maximal enhancement at a resonant pump frequency. On the other hand, circularly polarized pumps suppress the rCDW uniformly. Since the pumps break the time-reversal symmetry, they further trigger the emergent iCDW. Our work establishes  ultrafast optical pumping as an effective means to control the charge orders in the kagome metals $A$V$_3$Sb$_5$.

\begin{figure}[t]
\centering
\includegraphics[scale=1]{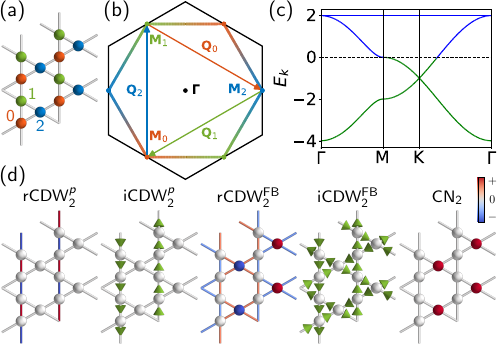}
\caption{\label{fig:kagome} Kagome lattice and charge orders. (a) The lattice structure with the sublattices labeled and colored. (b) The Brillouin zone, Fermi surface, high-symmetry points, and nesting vectors. The color along the Fermi surface indicates the sublattice weights, and the other objects are colored according to their indices. (c) The band structure. The occupied and unoccupied band segments are green and blue, respectively, and are separated by the dashed Fermi level. (d) Relevant charge orders in the post-pump dynamics. The site and bond colors indicate the density deviations from average (see the color bar), while the bond arrows represent the currents. Here we show the $\alpha=2$ orders, and the $\alpha=0,1$ ones are related by $\text{C}_3$ rotations.}
\end{figure}

\textit{Kagome lattice and TrH rCDW.---}We study the repulsive Hubbard model of spinless fermions on the kagome lattice [Fig.~\ref{fig:kagome}(a)]
\begin{equation}
\label{eq:ham}
H=-t_1\sum_{\epvl{ij}}c_i^\dagger c_j+\frac{1}{2}U_1\sum_{\epvl{ij}}c_i^\dagger c_j^\dagger c_jc_i.
\end{equation}
Here $c_i^{(\dagger)}$ annihilates (creates) a fermion at a lattice site $i$. The noninteracting Hamiltonian involves the nearest-neighbor hopping $t_1=1$. Setting the fermion filling at $n_\text{f}=5/12$, the Fermi level lies at the ``pure''-type Van Hove singularity ($p$VHS) in the middle band [Figs.~\ref{fig:kagome}(b) and \ref{fig:kagome}(c)]. The hexagonal Fermi surface connects the $p$VHS points $\mbf M_{a=0,1,2}$, whose parallel edges $\text{FS}_a\equiv\mbf M_{a+1}\text{-}\mbf M_{a+2}$ ($3\equiv0$) support strong nesting at three momenta $\mbf Q_{0,1,2}\equiv\mbf M_{0,1,2}$. Under the nearest-neighbor repulsion $U_1=1$, our Hartree-Fock theory \cite{lin24prb} (Supplemental Material, SM, Sec.~I \cite{supp}) obtains the mean-field ground state $\ket{\Psi_\text{GS}}$ with a $3Q$ rCDW (Fig.~\ref{fig:feature}). This ground state is $\text{C}_{6v}$-symmetric, where the rCDW strengths at $\mbf M_{0,1,2}$ are uniform. The rCDW at $\mbf M_a$ manifests the strongest particle-hole condensates $\text{Re}[\epvl{(c^\text{$p$VHS}_{\mbf M_{a+1}})^\dagger c^\text{$p$VHS}_{\mbf M_{a+2}}}]$ from the $p$VHS (rCDW$^p$). Since the sublattice polarization assigns each $\mbf M_a$ to a single sublattice $a$ \cite{kiesel12prb,wu23prb,lin24ax}, the $3Q$ rCDW$^p$ manifests the primary $2\times2$ TrH charge bond order \cite{lin21prb,park21prb,christensen21prb}. The $3Q$ rCDWs also involve the other two bands at the mean-field level, including the bottom ``mixed''-type VHS and the top flat band (rCDW$^{m,\text{FB}}$). These rCDWs have weaker strengths, approximately half that of the rCDW$^p$.

To draw a connection to the experiments \cite{ratcliff21prm,wang21prb,yu23prb,wu22prb,azoury23pnas,zhong24ax,ning24ax,xing24n}, we estimate parameters (energy and space scales) of the Hubbard model (\ref{eq:ham}) to describe the kagome metals $A$V$_3$Sb$_5$ (SM Sec.~II\,A \cite{supp}). For the energy scale, we use a reasonable magnitude $1\text{ energy unit}\equiv1\text{ eV}$ for the band structure \cite{wilson24nrm}. Meanwhile, $1\text{ length unit}\equiv2.75\times10^{-10}\text{ m}$ is adopted by matching the unit site distance with CsV$_3$Sb$_5$ \cite{tsirlin22sp}.

\textit{Pump pulse and dynamics.---}Our main focus is the dynamics of the ground state $\ket{\Psi(t=0)}=\ket{\Psi_\text{GS}}\rightarrow\ket{\Psi(t)}$ under a pump pulse. Here the time $t$ is assigned a scale $1\text{ time unit}\equiv0.659\text{ fs}$, and the corresponding frequency scale is $1 \text{ frequency unit}\equiv241.7\text{ THz}$ (SM Sec.~II\,A \cite{supp}). A pump pulse can be modeled by a time-dependent gauge field \cite{shao21prb,yang24ax}
\begin{equation}
\mbf A(t)=A_\text{c}e^{-(t-t_\text{c})^2/2\sigma_\text{t}^2}\mbf e_\text{p}(t,\omega_\text{c}).
\end{equation}
We fix the center time $t_\text{c}=25$ and the characteristic peak width $\sigma_\text{t}=3$, which give an approximate pulse duration $t\in[0,50]$ with $\exp[-(t-t_\text{c})/2\sigma_\text{t}^2]>10^{-15}$. The corresponding duration $32.9\text{ fs}$ is comparable to that in experiments \cite{ratcliff21prm,wang21prb,yu23prb,wu22prb,azoury23pnas,zhong24ax,ning24ax}. Meanwhile, the tunable parameters include the center amplitude $A_\text{c}$ and frequency $\omega_\text{c}$, as well as the oscillating polarization vector $\mbf e_\text{p}(t,\omega_\text{c})$. With the parameter ranges $\omega_\text{c}\in[1,3]$ and $A_\text{c}\in[0.02,0.1]$, the pump pulse carries the energy $\omega_\text{c}\text{ eV}$ and the fluence $F=0.001\text{-}0.6$ $\text{mJ}/\text{cm}^2$ (SM Sec.~II\,B \cite{supp}). Both of these values are consistent with the experiments \cite{ratcliff21prm,wang21prb,yu23prb,wu22prb,azoury23pnas,zhong24ax,ning24ax}. Through the Peierls substitution \cite{chou17prb,delatorre21rmp,shao21prb,yang24ax}
\begin{equation}
-t_1\sum_{\epvl{ij}}c_i^\dagger c_j\rightarrow-t_1\sum_{\epvl{ij}}c_i^\dagger e^{-i\mbf A(t)\cdot(\mbf r_i-\mbf r_j)}c_j,
\end{equation}
the pump pulse makes the Hamiltonian $H(t)$ time-dependent during the short pulse duration. We treat the system as spatially uniform, meaning we focus on the laser-illuminated spots.

The pump pulse can drive nonequilibrium dynamics through various mechanisms \cite{orenstein12pt,giannetti16aip,delatorre21rmp,boschini24rmp}. In addition to the variations in the TrH order, the dynamics also accommodate emergent charge orders with nonzero time averages. Here we discuss three scenarios which are relevant to our study. First, the pump pulse shakes the band structure $\epsilon_{\mbf k}(t)=\epsilon_{\mbf k+\mbf A(t)}$ and drags the fermion distribution away from the Fermi-Dirac one \cite{giannetti16aip,delatorre21rmp}. This nonequilibrium distribution modifies the free-energy landscape, thereby driving nontrivial dynamics of the orders. In our model, the shaking reduces the nesting singularity along $\text{FS}_a$ and suppresses the corresponding rCDW$^p$ at $\mbf M_a$. With the dominant rCDW suppressed, the other competing charge orders can emerge under the dynamics. Second, the pump pulse can break the symmetries and triggers the corresponding orders. In particular, the breaking of rotation and time-reversal symmetries can determine the relevant charge orders in the post-pump dynamics. Third, the pump pulse can induce a Floquet-like coupling \cite{delatorre21rmp} and couples the occupied bands to the unoccupied flat band. This effect can enhance the charge orders in the flat band, such as the rCDW$^\text{FB}$.

Employing the time-dependent Hartree-Fock theory (SM Sec.~III \cite{supp}), we study the post-pump dynamics of the state $\ket{\Psi(t)}$ under the Hamiltonian $H(t)$. Our time evolution runs up to $t=500$, which converts to $0.329\text{ ps}$. To analyze the dynamics, we compute the representative charge orders $\Delta_{\alpha=0,1,2}$ along the time evolution. Since the state $\ket{\Psi(t)}$ remains $2\times2$ periodic in the regimes we consider, the relevant charge orders narrow down to the $2\times1$ and $1\times1$ ones [Fig.~\ref{fig:kagome}(e)] (SM Sec.~IV\,A \cite{supp}). For the $2\times1$ orders at $\mbf M_{\alpha=0,1,2}$, we consider the r and iCDWs. The rCDW$^p$ is the primary charge order under the time evolution. Meanwhile, the iCDW$^p$ is relevant under the Fermi-surface nesting \cite{lin19prb,lin21prb,park21prb,dong23prb,fu24ax}. Note that the r and iCDW$^\text{FB}$ can be enhanced under the pump pulse, while their $m$VHS counterparts remain relatively weak. On the other hand, the $1\times1$ orders at $\boldsymbol\Gamma$ are included as distinct site and bond orders. In particular, the sublattice polarization promotes the CN \cite{kiesel13prl,wang13prb,profe24ax,lin24ax,fu24ax}, where the charge density becomes imbalanced among the three sublattices $\alpha=0,1,2$.

An important feature of the post-pump dynamics is the strong oscillations. Such oscillations exhibit two major bundles in the frequency spectrum (SM Sec.~IV\,B \cite{supp}). At high frequencies $4\lesssim\omega\lesssim6$, a major bundle captures the rapid breaking and recombination of the particle-hole condensates at $O(10^3)\text{ THz}$. The other major bundle forms at low frequencies $0\leq\omega\lesssim1$ and contains the coherent oscillations of the charge-order collective modes at $10\text{-}200\text{ THz}$. Since the collective-mode oscillations occur in the free-energy valleys \cite{giannetti16aip,delatorre21rmp}, the peak frequencies correspond to their effective masses. Our analysis focuses on the collective-mode oscillations in the low-frequency bundle. To remove the messy oscillations at high frequencies $\omega>2\pi/10$, we smear the data with a time average $\bar\Delta_\alpha(t)=(1/10)\int_{t-10}^tdt'\Delta_\alpha(t')$ at each time $t$. This treatment is suitable for a comparison with the experiments, whose temporal resolutions are usually above $5\text{ fs}$.

Remarkably, the dynamical phenomena depend strongly on the pump pulse. In the following, we present our analyses of the post-pump dynamics under the linearly and circularly polarized pulses (SM Sec.~IV\,C \cite{supp}).

\begin{figure}[t]
\centering
\includegraphics[scale=1]{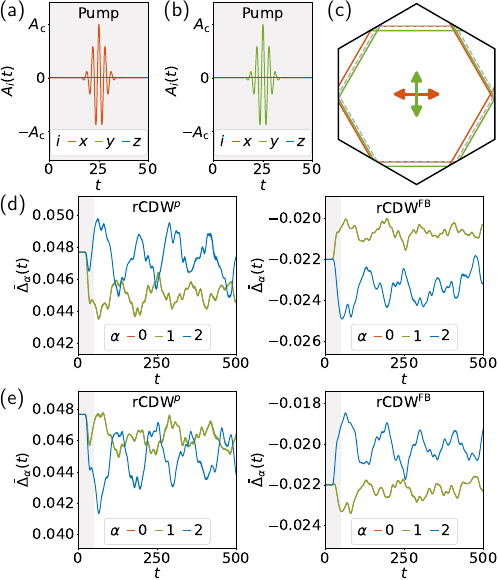}
\caption{\label{fig:linrcdw} Directional preference of the rCDWs under linearly polarized pumps. The polarizations are set along $\mbf e_{\text{p},0}=$ (a) $(1,0,0)$ and (b) $(0,1,0)$ with nonzero (a) $x$ and (b) $y$ component. (c) The Fermi surface is shaken in the corresponding directions. (d)(e) Directional preferences of the rCDW$^{p,\text{FB}}$ under the two types of pumps with $\omega_\text{c}=2$ and $A_\text{c}=0.06$. The $\alpha=0,1$ orders evolve identically in each case, while the $\alpha=2$ order undergoes a different evolution. Note that the rCDW$^\text{FB}$ can be enhanced in the post-pump dynamics. The relevant time scales under our unit conversion are pulse duration (gray time domain) $t=50\equiv32.9\text{ fs}$ and total time $t=500\equiv0.329\text{ ps}$.}
\end{figure}

\textit{Linearly polarized pump.---} We first consider linearly polarized pumps, which are defined by the oscillating polarization vector $\mbf e_\text{p}(t,\omega_\text{c})=\cos[\omega_\text{c}(t-t_\text{c})]\mbf e_{\text{p},0}$ \cite{yang24ax}. The fixed unit vector $\mbf e_{\text{p},0}$ sets the polarization direction. We adopt two important polarization directions $\mbf e_{\text{p},0}=(1,0,0)$ and $(0,1,0)$, which are parallel and perpendicular to $\mbf M_2$, respectively [Figs.~\ref{fig:linrcdw}(a) and \ref{fig:linrcdw}(b)]. The pump pulses break the $\text{C}_{6v}$ symmetry down to $\text{C}_{2v}$, where the reflection symmetries remain along $\mbf M_2$ and $\mbf Q_2$. Correspondingly, a splitting is expected between rCDW$^p_{0,1}$ and rCDW$^p_2$. Under the $(1,0,0)$ pump, $\text{FS}_{0,1}$ are shaken away from their original lines, while $\text{FS}_2$ slides along itself [Figs.~\ref{fig:linrcdw}(c)]. This difference implies a stronger suppression of rCDW$^p_{0,1}$ than rCDW$^p_2$. Our analysis confirms the preference for rCDW$^p_2$ under the time evolution [Figs.~\ref{fig:linrcdw}(d)]. Notably, this directional preference of the rCDW is consistent with the laser-coupled-STM experiment on RbV$_3$Sb$_5$ \cite{xing24n}. The directional preference also occurs to the rCDW$^\text{FB}$. Interestingly, rCDW$^\text{FB}_2$ is completely enhanced under the Floquet-like coupling of the pump pulse. On the other hand, the $(0,1,0)$ pump drives the opposite directional preference of the rCDW [Figs.~\ref{fig:linrcdw}(e)]. This opposite behavior results from the farther shake of $\text{FS}_2$ than $\text{FS}_{0,1}$ [Figs.~\ref{fig:linrcdw}(c)]. Note that this directional preference experiences a faster decay than under the $(1,0,0)$ pump, since the $1Q$ orders are energetically more favorable than the $2Q$ ones \cite{park21prb,lin22prb}.

\begin{figure}[t]
\centering
\includegraphics[scale=1]{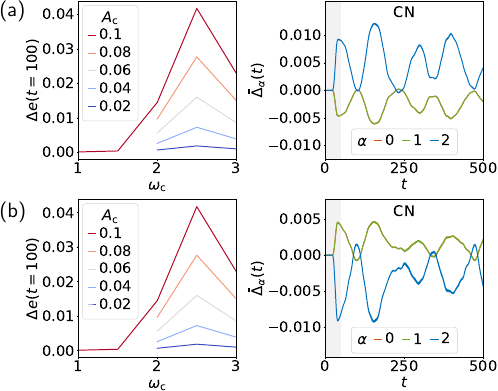}
\caption{\label{fig:lincn} Emergent CN under the linearly polarized pumps. The resonance peaks appear at $\omega_\text{c}\approx2.5$ in the injected energy $\Delta e(t=100)$ for both $\mbf e_{\text{p},0}=$ (a) $(1,0,0)$ and (b) $(0,1,0)$. The emergent CN is strongly enhanced at $\omega_\text{c}=2.5$ with $A_\text{c}=0.06$. The $\alpha=0,1$ orders evolve identically in each case, while the $\alpha=2$ order undergoes a different evolution. The relevant time scales under our unit conversion are pulse duration (gray time domain) $t=50\equiv32.9\text{ fs}$ and total time $t=500\equiv0.329\text{ ps}$.}
\end{figure}

We further inspect the dependence of the post-pump dynamics on the center frequency $\omega_\text{c}$. Intuitively, one expects more site-averaged injected energy $\Delta e(t)=e(t)-e(0)$ from a higher-frequency pump. However, we find a surprising peak at $\omega_\text{c}\approx2.5$ across all center amplitudes $A_\text{c}\in[0.02,0.1]$ (Fig.~\ref{fig:lincn}). Remarkably, the emergent CN is maximally enhanced at this peak, accompanied by the maximal suppression of the rCDW$^p$. These phenomena can be understood from the resonance in a Rabi-like oscillation \cite{shao21prb}. A Rabi oscillation occurs in a two-level system $H_\text{R}(t)=(\epsilon'/2)\sigma_3+g'\cos(\omega't)\sigma_1$, where $\sigma_{1,3}$ are the Pauli matrices and $2\epsilon'$ is the energy gap. When the ground state $(0,1)$ at $t=0$ experiences the periodic drive at amplitude $g'$ and frequency $\omega'$, it oscillates at a Rabi frequency $\omega_\text{R}=[(\omega'-\epsilon')^2+g'^2]^{1/2}/2$. In our model, the pump pulse drives a Rabi-like oscillation between the ground-state rCDW$^p$ and the excited-state CN in the short pulse duration. As the center frequency reaches the resonant value $\omega_\text{c}\approx2.5$, the driven oscillation terminates with a strong weight in the CN. This resonance allows the state $\ket{\Psi(t)}$ to absorb maximal injected energy $\Delta e(t)$ and develop an enhanced CN. Our model contains a complex intertwining of several charge orders, which goes far beyond the simple two-level model. Therefore, the Rabi-like oscillation does not convert the state $\ket{\Psi(t)}$ entirely to the CN. Note that the CN decays faster under the $(0,1,0)$ pump than under the $(1,0,0)$ one, since the 1-higher-2-lower CN is energetically more favorable than the 2-higher-1-lower one \cite{fu24ax,lin24ax}.

\begin{figure}[t]
\centering
\includegraphics[scale=1]{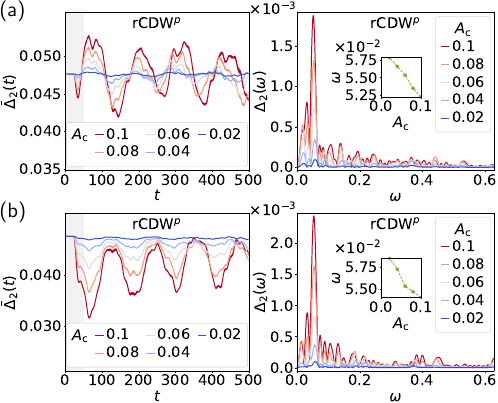}
\caption{\label{fig:linsoftening} Collective-mode softening of rCDW$^p$ under the linearly polarized pumps. The softening occurs for both $\mbf e_{\text{p},0}=$ (a) $(1,0,0)$ and (b) $(0,1,0)$, as can be observed in the time evolution and the frequency spectrum with $\omega_\text{c}=2$. (Inset) The major peak shifts toward lower frequency with increasing center amplitude $A_\text{c}$. The relevant time and frequency scales under our unit conversion are pulse duration (gray time domain) $t=50\equiv32.9\text{ fs}$, total time $t=500\equiv0.329\text{ ps}$, and major-peak frequency scale $\omega=0.05\equiv12.085\text{ THz}$.}
\end{figure}

We also investigate the dependence of the post-pump dynamics on the center amplitude $A_\text{c}$. Notably, the collective modes experience the pump-induced softening, where the major oscillation slows down with increasing center amplitude $A_\text{c}$ (Fig.~\ref{fig:linsoftening}). This trend is further confirmed in the frequency spectrum, where the major peak shifts toward lower frequency. The collective-mode softening can be understood from the free-energy landscape. Under a stronger pump pulse, the depth of a free-energy valley decreases more. Therefore, the effective mass and the corresponding oscillation frequency decrease with increasing center amplitude $A_\text{c}$.

\begin{figure}[t]
\centering
\includegraphics[scale=1]{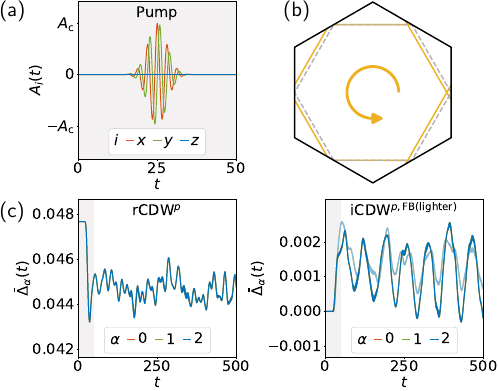}
\caption{\label{fig:cir} Dynamics under the circularly polarized pump. (a) The pump pulse drives (b) a circular shake of the Fermi surface. (c) The uniformly suppressed rCDW$^p$, as well as the emergent iCDW$^p$ and iCDW$^\text{FB}$ (lighter color), under the pump with $\omega_\text{c}=2$ and $A_\text{c}=0.06$. In each case, the smeared time evolution of the $\alpha=0,1,2$ orders are nearly identical. The relevant time scales under our unit conversion are pulse duration (gray time domain) $t=50\equiv32.9\text{ fs}$ and total time $t=500\equiv0.329\text{ ps}$.}
\end{figure}

\textit{Circularly polarized pump.---}Our next subject is the circularly polarized pump with a rotating unit polarization vector $\mbf e_\text{p}(t,\omega_\text{c})=(\cos[\omega_\text{c}(t-t_\text{c})],\sin[\omega_\text{c}(t-t_\text{c})],0)$ [Fig.~\ref{fig:cir}(a)] \cite{shao21prb}. Since the pump pulse breaks the $\text{C}_{6v}$ symmetry circularly, the resulting dynamics exhibits the same symmetry breaking at any time $t$. Under the circular shaking of $\text{FS}_{0,1,2}$ [Fig.~\ref{fig:cir}(b)], the rCDW$^p_{0,1,2}$ experience cyclic oscillations in the high-frequency bundle. Nevertheless, the low-frequency dynamics remain nearly $\text{C}_{6v}$-symmetric, as long as the center frequency is high enough $\omega_\text{c}\gtrsim1$ for at least a complete cycle within the characteristic peak width $[t_\text{c}-\sigma_\text{t},t_\text{c}+\sigma_\text{t}]$. Indeed, our analysis shows a uniform suppression of all three rCDW$^p_{0,1,2}$ in the post-pump dynamics [Fig.~\ref{fig:cir}(c)]. Importantly, the circularly polarized pump breaks the time-reversal symmetry and triggers the $3Q$ iCDW$^p$. Although the emergent CLCs are too weak to induce a Chern insulator \cite{venderbos16prb,lin19prb,lin21prb}, they still induce nontrivial Berry curvatures with possible experimental manifestations. Interestingly, the $3Q$ iCDW$^\text{FB}$ also emerges, which can even surpass the $3Q$ iCDW$^p$. A resonance peak is again observed at $\omega_\text{c}\approx2.5$, together with an enhancement of the $3Q$ iCDWs. Note that an opposite center frequency $\omega_\text{c}\rightarrow-\omega_\text{c}$ flips the signs of the iCDWs.

Compared to the CNs under the linearly polarized pumps, the iCDWs under the circularly polarized pump appear weaker and less stable. This weakness can be understood from the free energy of the complex CDW$^p_{0,1,2}$ with $\Delta_{\alpha}=|\Delta_\alpha|\exp(i\phi_\alpha)$ \cite{lin21prb}. At the cubic order, the term $-|\Delta_1||\Delta_2||\Delta_3|\cos(\phi_1+\phi_2+\phi_3)$ imposes a total phase condition $\phi_1+\phi_2+\phi_3=2\pi$ for the energetic favor. The $3Q$ rCDW$^p$ carries the phases $\phi_{0,1,2}=0$ and satisfies the condition. To accommodate the iCDW$^p_{0,1,2}$, the isotropic solutions $\phi_{0,1,2}=\pm2\pi/3$ both require negative rCDW$^p_{0,1,2}$. The pump pulse is not strong enough to overcome the free-energy barrier of this sign change. Therefore, it only drives the complex CDW$^p$ to a relatively unstable point, which decays rapidly in time.

\textit{Discussion and outlook.---} We have analyzed ultrafast optical control of the various charge orders in the kagome metals $A$V$_3$Sb$_5$. There remains enormous uncharted territory to explore along this research frontier. First, while we have explained the observed dynamical phenomena intuitively in terms of oscillations in an effective free energy landscape, a theory yielding that landscape directly could provide more precise characterizations. The establishment of such a theory is an important direction for future work. Second, the time-dependent Hartree-Fock theory is sufficiently powerful to capture our main focus: the ultrafast optical control of  charge orders. However, it does not include all of the possible relaxation processes in the late-time dynamics. Future works with numerically more expensive methods could examine the late-time dynamics in our results more accurately. Third, our work aims at identifying the most important dynamical phenomena by adopting a representative simple electronic model. There are additional properties in the kagome metals $A$V$_3$Sb$_5$ that may lead to quantitative differences in the post-pump dynamics. For example, the electron-phonon coupling may drag the collective-mode oscillations and reduce the frequency below $10\text{ THz}$ \cite{ratcliff21prm,wang21prb,yu23prb,wu22prb,azoury23pnas,zhong24ax,ning24ax}. Moreover, the V-Sb bonds, either in-plane or out-of-plane, may modify the Peierls substitution. Interestingly, the three-dimensional structure allows the $(0,0,1)$ linearly polarized pump to be effective, which shakes the Fermi surface in the out-of-plane direction. The in-plane correspondence is an expansion-contraction shaking, similar to a chemical-potential shaking \cite{yu23prb}. Since the shaking does not break the $\text{C}_{6v}$ symmetry, a uniform suppression is expected for all three rCDW$^p_{0,1,2}$.

Our results can be directly examined in  experiments. With the $(1,0,0)$ linearly polarized pump parallel to $\mbf M_2$, the directional preference of the rCDW$^p$ has been observed in the laser-coupled STM \cite{xing24n}. An equivalent probe with the perpendicular $(0,1,0)$ pump may be interesting for future experiments. The search for the emergent CN is also feasible by tuning the laser frequency. Note that these real charge orders may also be probed by other methods, such as trXRD. Meanwhile, the $3Q$ iCDW$^p$ under the circularly polarized pump could show up as nontrivial Kerr effects in time-resolved optical polarization rotation \cite{wu22prb}. On the other hand, our results may also apply to the magnetic kagome material FeGe, in which a similar CDW develops at around $100$ K \cite{teng22n,lin24prb,linzhou24ax}.
Finally, our work fits within the broader context of ultrafast optical control in quantum materials with multiple correlated phases, and further investigations into other materials with competing charge, spin, and superconducting \cite{claassen19np} orders can build on the approach implemented here.

\begin{acknowledgments}
The authors thank Joseph Orenstein for fruitful discussions. This work was supported by the MURI program of the Air Force Office of Scientific Research under Grant No. FA9550-22-1-0270 and the U.S. Department of Energy Office of Science National Quantum Information Science Research Centers as part of the Q-NEXT center. Y.-P.L. acknowledges fellowship support from the Gordon and Betty Moore Foundation through the Emergent Phenomena in Quantum Systems (EPiQS) program. V.M was supported by the Gordon and Betty Moore Foundation's EPiQS initiative through Grant No. GBMF9465. Parts of the numerical computations were performed on the Lawrencium cluster at Lawrence Berkeley National Laboratory.
\end{acknowledgments}




\bibliography{reference}

\clearpage
\onecolumngrid

\begin{center}{\large\bf
Supplemental Material for\\``Ultrafast optical control of charge orders in kagome metals''
}\end{center}

\setcounter{secnumdepth}{3}
\setcounter{section}{0}
\setcounter{equation}{0}
\setcounter{figure}{0}
\renewcommand{\theequation}{S\arabic{equation}}
\renewcommand{\thefigure}{S\arabic{figure}}
\newcommand\Scite[1]{[S\citealp{#1}]}
\makeatletter \renewcommand\@biblabel[1]{[S#1]} \makeatother

\addtocontents{toc}{\string\tocdepth@restore}

\tableofcontents

\section{Hartree-Fock theory}

In this section, we provide a comprehensive introduction to the Hartree-Fock theory and its numerical scheme. The specific implementation in our work is also explained.

\subsection{General formalism}

We begin with an introduction to the general formalism of the Hartree-Fock theory \Scite{bultinck20prx,lin24prb}. Consider a general model of the interacting-fermion systems
\begin{equation}
\label{suppeq:model0}
H=\sum_{ab}\mathcal H_{0,ab}c_a^\dagger c_b+\frac{1}{2}\sum_{abcd}U_{acdb}c_a^\dagger c_c^\dagger c_dc_b.
\end{equation}
The noninteracting Hamiltonian $\mathcal H_{0,ab}$ contains the kinetic energy and the background potential. Meanwhile, the interacting term $U_{acdb}$ defines the interaction. The fermionic annihilation and creation operators, $c_a$ and $c_a^\dagger$, can be defined in any basis, such as position, momentum, eigenstate, etc. Assume that the system has $N_\text{DOF}$ degrees of freedom. With a properly chosen filling $n_\text{f}$, the system hosts an integer number $N=n_\text{f}N_\text{DOF}$ of fermions. Our focus lies in the interacting states $\ket{\Psi}$ of these $N$ fermions, especially the ground states $\ket{\Psi_\text{GS}}$, at zero temperature. At the mean-field level, we assume that these states take the Slater-determinant forms, where $N$ effectively noninteracting states $\psi_\alpha$ are occupied under the Fermi statistics.

Given a state $\ket{\Psi}$, we can define its expectation values $\epvl{\cdots}=\braket{\Psi}{\cdots}{\Psi}$. In particular, the density matrix $P_{ab}=\epvl{c_b^\dagger c_a}$, also known as the propagator or the two-point Green's function, serves as a powerful representation of the state. At the mean-field level, the interaction effects only show up in the particle-hole $\epvl{c^\dagger c}$ and particle-particle $\epvl{(cc)^{(\dagger)}}$ condensates. While the later is not considered in this work, the former is perfectly captured by the density matrix $P$. An important property of the density matrix is the projector condition $P^2=P$. This condition can be easily observed in the diagonal basis $P=\text{diag}(1,1,\dots,1,0,0,\dots)$ under the $N$-state occupation. Alternatively, we can understand it by recognizing the matrix multiplication as a path integral within the $N$-electron subspace. On the other hand, the energy of a state $\ket{\Psi}$ takes the form
\begin{equation}
\begin{aligned}
E[P]&=\epvl{H}
=\sum_{ab}\mathcal H_{0,ab}\epvl{c_a^\dagger c_b}+\frac{1}{2}\sum_{abcd}U_{acdb}\epvl{c_a^\dagger c_c^\dagger c_dc_b}.
\end{aligned}
\end{equation}
The first term can be directly rewritten in terms of the density matrix
\begin{equation}
\sum_{ab}\mathcal H_{0,ab}\epvl{c_a^\dagger c_b}
=\sum_{ab}\mathcal H_{0,ab}P_{ba}
=\Tr(\mathcal H_0P).
\end{equation}
Meanwhile, the second term is solved by the Wick's theorem
\begin{equation}
\label{suppeq:wick}
\epvl{c_a^\dagger c_c^\dagger c_dc_b}=
\contraction[2ex]{}{c_a^\dagger}{c_c^\dagger c_d}{c_b}\contraction{c_a^\dagger}{c_c^\dagger}{}{c_d}\epvl{c_a^\dagger c_c^\dagger c_dc_b}+\contraction[2ex]{}{c_a^\dagger}{c_c^\dagger}{c_d}\contraction{c_a^\dagger}{c_c^\dagger}{c_d}{c_b}\epvl{c_a^\dagger c_c^\dagger c_dc_b}
=\epvl{c_a^\dagger c_b}\epvl{c_c^\dagger c_d}-\epvl{c_a^\dagger c_d}\epvl{c_c^\dagger c_b}
=P_{ba}P_{dc}-P_{da}P_{bc},
\end{equation}
leading to
\begin{equation}
\frac{1}{2}\sum_{abcd}U_{acdb}\epvl{c_a^\dagger c_c^\dagger c_dc_b}
=\frac{1}{2}\sum_{abcd}U_{acdb}(P_{ba}P_{dc}-P_{da}P_{bc})
=\frac{1}{2}\sum_{abcd}(U_{acdb}-U_{acbd})P_{ba}P_{dc}.
\end{equation}
The energy is thus derived as
\begin{equation}
\begin{aligned}
E[P]=\Tr(\mathcal H_0P)+\frac{1}{2}\sum_{abcd}(U_{acdb}-U_{acbd})P_{ba}P_{dc}.
\end{aligned}
\end{equation}

We want to find the ground state $\ket{\Psi_\text{GS}}$ with the minimal energy $E[P_\text{GS}]$. The minimization is conducted by a variation with respect to the density matrix $P$. Importantly, the variation of the energy
\begin{equation}
\frac{\delta E[P]}{\delta P}=\mathcal H_\text{HF}[P]
\end{equation}
gives the Hartree-Fock Hamiltonian
\begin{equation}
\label{suppeq:hfham}
\mathcal H_\text{HF}[P]_{ab}=\mathcal H_{0,ab}+\mathcal V_\text{HF}[P]_{ab}=\mathcal H_{0,ab}+\sum_{cd}(U_{acdb}-U_{acbd})P_{dc}.
\end{equation}
This Hamiltonian describes an effectively noninteracting theory at the mean-field level
\begin{equation}
\label{suppeq:hfmft}
H_\text{HF}=\sum_{ab}\mathcal H_{\text{HF},ab}c_a^\dagger c_b,
\end{equation}
where the interaction effects contribute to two distinct terms in the Hartree-Fock potential $\mathcal V_\text{HF}[P]$. The first term is a simple density correction to the background potential, known as the Hartree potential. Meanwhile, the second term results from the exchange under the Fermi statistics, often named as the Fock or exchange potential. The energy can be written in terms of the Hartree-Fock Hamiltonian as
\begin{equation}
E[P]=\frac{1}{2}\Tr(P(\mathcal H_0+\mathcal H_\text{HF}[P])).
\end{equation}
When we minimize the energy, it is important to incorporate the projector condition $P^2=P$. This treatment is achieved by considering the cost function
\begin{equation}
C[P]=E[P]-\Tr(X(P^2-P))
\end{equation}
with a Lagrange multiplier $X$. The variation of the cost function gives the minimization condition
\begin{equation}
0=\left.\frac{\delta C[P]}{\delta P}\right|_{P=P_\text{GS}}=\mathcal H_\text{HF}[P_\text{GS}]-(P_\text{GS}X+XP_\text{GS}-X).
\end{equation}
Subtracting the left and right multiplications with $P_\text{GS}$, we find the self-consistent equation in a commutator form
\begin{equation}
[P_\text{GS},\mathcal H_\text{HF}[P_\text{GS}]]=0.
\end{equation}
This commutator equation implies a self-consistent density-matrix condition: In the ground state $\ket{\Psi_\text{GS}}$, the density matrix $P_\text{GS}$ should be equal to the one formed by the $N$ lowest-lying eigenstates $\psi_{\text{GS},\alpha}$ of its corresponding Hartree-Fock Hamiltonian $H_\text{HF}[P_\text{GS}]$
\begin{equation}
\label{suppeq:selfcons}
P_\text{GS}=\sum_{\alpha=1}^N\psi_{\text{GS},\alpha}\psi_{\text{GS},\alpha}^*.
\end{equation}

\subsection{Numerical scheme}

We have explored the general formalism of the Hartree-Fock theory. The next step is to implement it numerically on the systems of our interest. While there exists a few different numerical schemes, the one we use is based on an iterative implementation of the self-consistent density-matrix condition (\ref{suppeq:selfcons}). The input to this iterative algorithm is an initial density matrix $P_0$, which is usually randomized from $N$ orthonormal states $\psi_\alpha$. Specific forms of $P_0$ can be chosen in order to achieve certain results. For the initial iteration $m=0$, the Hartree-Fock Hamiltonian $\mathcal H_{\text{HF},m}=\mathcal H_\text{HF}[P_m]$ and the energy $E_m=(1/2)\Tr(P_m(\mathcal H_0+\mathcal H_{\text{HF},m}))$ are computed from $P_m=P_0$. (Optional) If the optimal damping algorithm (ODA) \Scite{kudin02jcp} is applied to accelerate the convergence, the additional density matrix $\tilde P_m=P_m$ and Hartree-Fock Hamiltonian $\tilde{\mathcal H}_{\text{HF},m}=\mathcal H_{\text{HF}}[\tilde P_m]$ are set up for the initial iteration $m=0$.

In the $m$-th iteration, the variational update proceeds as follows:
\begin{enumerate}
\item Diagonalize the Hartree-Fock Hamiltonian $\mathcal H_{\text{HF},m}=U_\epsilon D_\epsilon U_\epsilon^\dagger$, where $D_\epsilon=\text{diag}(\epsilon_1,\epsilon_2,\dots)$ with $\epsilon_1<\epsilon_2<\dots$ are the eigenvalues and $U_\epsilon=(\psi_1,\psi_2,\dots)$ are the eigenstates of $\mathcal H_{\text{HF},m}$. (Optional) If the ODA is applied, diagonalize $\tilde{\mathcal H}_{\text{HF},m}$ instead.
\item Assemble the new density matrix $P_{m+1}=U_\epsilon D_NU_\epsilon^\dagger$, where $D_N=\text{diag}(1,1,\dots,1,0,0,\dots,0)$ selects the $N$ lowest-lying eigenstates.
\item Compose the new Hartree-Fock Hamiltonian $\mathcal H_{\text{HF},m+1}=\mathcal H_\text{HF}[P_{m+1}]$.
\item Compute the new energy $e_{m+1}=E_{m+1}/N_\text{DOF}$.
\item Compute the desired physical observables.
\item Check the convergence: Stop the iteration if the variations in the energy $|e_{m+1}-e_m|<\delta e$ and the density-matrix elements $|P_{m+1,ab}-P_{m,ab}|<\delta p$ are small enough. The criteria $\delta e,\delta p=10^{-15}$ are chosen in our computation, although the density-matrix elements may only converge to $O(10^{-15})$ for large system sizes.
\item (Optional) Adopt the ODA \Scite{kudin02jcp}, which interpolates between $P_{m+1}$ and $\tilde P_m$ to minimize the energy.
\begin{enumerate}
\item[i.] Compute the variation of the density matrix $\delta P_m=P_{m+1}-\tilde P_m$ and the Hartree-Fock Hamiltonian $\delta \mathcal H_{\text{HF},m}=\mathcal H_{\text{HF},m+1}-\tilde{\mathcal H}_{\text{HF},m}$.
\item[ii.] Consider the energy of an interpolation $\tilde P_m+\lambda_m\delta P_m$ between $P_{m+1}$ and $\tilde P_m$ with $\lambda_m\in[0,1]$
\begin{equation}
\begin{aligned}
E[\tilde P_m+\lambda_m\delta P_m]
&=\frac{1}{2}\Tr((\tilde P_m+\lambda_m\delta P_m)(\mathcal H_0+\mathcal H_\text{HF}[\tilde P_m+\lambda_m\delta P_m]))\\
&=\frac{1}{2}\Tr((\tilde P_m+\lambda_m\delta P_m)(\mathcal H_0+\tilde H_{\text{HF},m}+\lambda_m\delta H_{\text{HF},m})).
\end{aligned}
\end{equation}
This energy is a quadratic function of the interpolation parameter $\lambda_m$
\begin{equation}
E[\tilde P_m+\lambda_m\delta P_m]=\tilde E_m+a_m\lambda_m+\frac{1}{2}b_m\lambda_m^2,
\end{equation}
where the prefactors are
\begin{equation}
\begin{aligned}
\tilde E_m&=\frac{1}{2}\Tr(\tilde P_m(\mathcal H_0+\tilde{\mathcal H}_{\text{HF},m})),\\
a_m&=\frac{1}{2}\Tr(\tilde P_m\delta H_{\text{HF},m}+\delta P_m(\mathcal H_0+\tilde H_{\text{HF},m})),\\
b_m&=\Tr(\delta P_m\delta H_{\text{HF},m}).
\end{aligned}
\end{equation}
Note that the linear terms satisfy the relation
\begin{equation}
\begin{aligned}
\Tr(\tilde P_m\delta H_{\text{HF},m})
&=\sum_{ab}\tilde P_{m,ba}\delta H_{\text{HF},m,ab}
=\sum_{ab}\tilde P_{m,ba}\sum_{cd}(U_{acdb}-U_{acbd})\delta P_{m,dc}\\
&=\sum_{cd}\delta P_{m,dc}\sum_{ab}(U_{cabd}-U_{cadb})\tilde P_{m,ba}
=\sum_{cd}\delta P_{m,dc}\tilde{\mathcal V}_{\text{HF},m,cd}\\
&=\Tr(\delta P_m\tilde{\mathcal V}_{\text{HF},m}),
\end{aligned}
\end{equation}
where the permutation rule of the interaction $U_{acdb}=U_{cabd}$ is applied between the first and second lines. This relation reduces the linear prefactor into
\begin{equation}
a_m=\Tr(\delta P_m\tilde{\mathcal H}_{\text{HF},m}).
\end{equation}
The point $\lambda_m=-a/b$ represents the extremum of the quadratic function. If the energy minimization is achieved at the intermediate point $\lambda_m\in(0,1)$, the corresponding interpolation is applied to the density matrix $\tilde P_{m+1}=\tilde P_m+\lambda_m\delta P_m$. Otherwise, the density matrix $\tilde P_{m+1}=P_{m+1}$ is chosen.
\item[iii.] Compose the Hartree-Fock Hamiltonian $\tilde H_{\text{HF},m+1}=H_\text{HF}[\tilde P_{m+1}]$.
\end{enumerate}
Note that the projector condition $P^2=P$ is broken under the ODA interpolation, which may raise some concerns in its eligibility. Nevertheless, the ODA interpolation usually occurs only in the early iterations to escape the local minima. Once the variation flow is guided onto an appropriate path toward the ground state, the later convergence proceeds just as usual.
\end{enumerate}
Note that the iterative algorithm does not guarantee the final state as the ground state with the minimal energy. Therefore, a sanity check is needed by examining the results across multiple random or assigned initial states.

\subsection{Implementation in this work}

We now discuss the specific implementation in our work.

\subsubsection{Real-space formalism}

To study the repulsive spinless fermions on the kagome lattice, we adopt a real-space formalism, where the fermionic basis is defined by the lattice sites $a=i\tau$. The indices $i$ and $\tau=0,1,2$ label the Bravais-lattice site and the three sublattices, respectively. The extended Hubbard model takes the form
\begin{equation}
H=\sum_{ii'\tau\tau'}\mathcal H_{0,ii'\tau\tau'}c_{i\tau}^\dagger c_{i'\tau'}+\frac{1}{2}\sum_{ii'\tau\tau'}U_{ii'\tau\tau'}c_{i\tau}^\dagger c_{i'\tau'}^\dagger c_{i'\tau'}c_{i\tau},
\end{equation}
with the nonzero elements $\mathcal H_{0,ii'\tau\tau'}=-t_1=-1$ and $U_{ii'\tau\tau'}=U_1=1$ for the nearest-neighbor pairs $\epvl{i\tau,i'\tau'}$. We choose the $18\times18$ Bravais lattice in our study, and the periodic boundary conditions are applied in both directions. The filling $n_\text{f}=5/12$ sets the Fermi level at the $p$-type Van Hove singularity in the middle band of the kagome lattice.

\subsubsection{Computation of charge orders}
\label{suppsssec:hfcharge}

To understand the ground states, we study the charge orders on the sites and bonds
\begin{equation}
P_{ii'\tau\tau'}=\epvl{c_{i'\tau'}^\dagger c_{i\tau}}.
\end{equation}
The site orders $P_{ii\tau\tau}$ are real, which characterize the onsite charge densities. Meanwhile, the bond orders $P_{ii'\tau\tau'}$ with $i\tau\neq i'\tau'$ are generally complex. For a specific bond, the real component of its bond order $\text{Re}(P_{ii'\tau\tau'})$ measures the charge-bond density. The imaginary component $\text{Im}(P_{ii'\tau\tau'})$ measures the charge-current density on this bond. As we will show in Sec.~\ref{suppsssec:current}, the charge current from $i\tau$ to a nearest neighbor $i'\tau'$ takes the form
\begin{equation}
\label{suppeq:current}
j_{ii'\tau\tau'}=-2t_1\text{Im}(P_{ii'\tau\tau'}).
\end{equation}

Given a density matrix, the symmetry breaking orders can be easily identified from its charge-order patterns. With the setup of our model, we find that the interaction-driven ground state is a $2\times2$ real charge-density wave (rCDW) at three nesting momenta $\mbf Q_{a=0,1,2}\equiv\mbf M_{a=0,1,2}$, usually known as the tri-hexagonal (TrH) charge-bond order (CBO).

\subsubsection{Band-structure analysis}

While the real-space formalism is already effective in understanding the interaction-driven ground states, further information can be gathered from the band-structure analysis in the momentum space. The band-structure analysis is based on a noninteracting model
\begin{equation}
\tilde H=\sum_{\tilde i\tilde i'\tilde\tau\tilde\tau'}\tilde{\mathcal H}_{\tilde i\tilde i'\tilde\tau\tilde\tau'}c_{\tilde i\tilde\tau}^\dagger c_{\tilde i'\tilde\tau'}.
\end{equation}
This model captures the noninteracting theory with the setup $\tilde{\mathcal H}=\mathcal H_0$ and $\tilde i\tilde\tau=i\tau$. For the study of interaction-driven ground states, we adopt the Hartree-Fock Hamiltonian $\tilde{\mathcal H}=\mathcal H_\text{HF}$, which represents the effectively noninteracting mean-field theory. Note that the ground states may break the translation symmetry and enlarge the periodicity. In this case, the Bravais lattice sites $\tilde i$ and the sublattices $\tilde\tau$ should be redefined with respect to the enlarged unit cells. This change also applies to the Brillouin zone in the momentum space. The Fourier transform to the momentum space is achieved by the basis transformation
\begin{equation}
c_{\tilde i\tilde\tau}
=\sum_{\mbf k}\innp{\tilde i\tilde\tau}{\mbf k\tilde\tau}c_{\mbf k\tilde\tau}
=\frac{1}{\tilde N_\text{BL}^{1/2}}\sum_{\mbf k}c_{\mbf k\tilde\tau}e^{i\mbf k\cdot\mbf r_{\tilde i\tilde\tau}}.
\end{equation}
Here $\tilde N_\text{BL}$ is the number of the Bravais-lattice sites $\tilde i$, and the plane-wave momentum eigenfunction $\innp{\tilde i\tilde\tau}{\mbf k\tilde\tau}=(1/\tilde N_\text{BL}^{1/2})\exp(i\mbf k\cdot\mbf r_{\tilde i\tilde\tau})$ with the lattice-site position $\mbf r_{\tilde i\tilde\tau}$ is adopted. Under the Fourier transform, the model $\tilde H$ becomes
\begin{equation}
\begin{aligned}
\tilde H
&=\sum_{\tilde i\tilde i'\tilde\tau\tilde\tau'}\tilde{\mathcal H}_{\tilde i\tilde i'\tilde\tau\tilde\tau'}\frac{1}{\tilde N_\text{BL}}\sum_{\mbf k\mbf k'}e^{i(-\mbf k\cdot\mbf r_{\tilde i\tilde\tau}+\mbf k'\cdot\mbf r_{\tilde i'\tilde\tau'})}c_{\mbf k\tilde\tau}^\dagger c_{\mbf k'\tilde\tau'}\\
&=\sum_{\mbf k\mbf k'}\sum_{\tilde i'\tilde\tau\tilde\tau'}\left[\frac{1}{\tilde N_\text{BL}}\sum_{\tilde i}e^{i(-\mbf k+\mbf k')\cdot\mbf r_{\tilde i\tilde\tau}}\right]\tilde{\mathcal H}_{\tilde i\tilde i'\tilde\tau\tilde\tau'}e^{-i\mbf k'\cdot(\mbf r_{\tilde i\tilde\tau}-\mbf r_{\tilde i'\tilde\tau'})}c_{\mbf k\tilde\tau}^\dagger c_{\mbf k'\tilde\tau'}.
\end{aligned}
\end{equation}
With the unity condition $\delta_{\mbf k\mbf k'}=(1/\tilde N_\text{BL})\sum_{\tilde i}\exp[i(-\mbf k+\mbf k')\cdot\mbf r_{\tilde i\tilde\tau}]$, the translation-symmetry condition $\tilde{\mathcal H}_{\tilde i\tilde i'\tilde\tau\tilde\tau'}=\tilde{\mathcal H}_{\tilde0(\tilde i'-\tilde i)\tilde\tau\tilde\tau'}$, and the translation $\tilde i'-\tilde i\rightarrow\tilde i'$, we have
\begin{equation}
\tilde H
=\sum_{\mbf k\mbf k'}\delta_{\mbf k\mbf k'}\sum_{\tilde i'\tilde\tau\tilde\tau'}\tilde{\mathcal H}_{\tilde0\tilde i'\tilde\tau\tilde\tau'}e^{-i\mbf k'\cdot(\mbf r_{\tilde0\tilde\tau}-\mbf r_{\tilde i'\tau'})}c_{\mbf k\tilde\tau}^\dagger c_{\mbf k'\tilde\tau'}
=\sum_{\mbf k}\left[\sum_{\tilde i'\tilde\tau\tilde\tau'}\tilde{\mathcal H}_{\tilde0\tilde i'\tilde\tau\tilde\tau'}e^{-i\mbf k\cdot(\mbf r_{\tilde0\tilde\tau}-\mbf r_{\tilde i'\tau'})}\right]c_{\mbf k\tilde\tau}^\dagger c_{\mbf k\tilde\tau'}.
\end{equation}
Defining the momentum-space Hamiltonian
\begin{equation}
\tilde{\mathcal H}_{\mbf k\tilde\tau\tilde\tau'}=\sum_{\tilde i'\tilde\tau\tilde\tau'}\tilde{\mathcal H}_{\tilde0\tilde i'\tilde\tau\tilde\tau'}e^{-i\mbf k\cdot(\mbf r_{\tilde0\tilde\tau}-\mbf r_{\tilde i'\tau'})},
\end{equation}
we arrive at the momentum-space representation of the model
\begin{equation}
\tilde H=\sum_{\mbf k}\tilde{\mathcal H}_{\mbf k\tilde\tau\tilde\tau'}c_{\mbf k\tilde\tau}^\dagger c_{\mbf k\tilde\tau'}.
\end{equation}
The band structure can be computed from the momentum-space Hamiltonian. Various important properties, including the gap structure, the Fermi surface, and the band topology, can be further determined.

\section{Comparison of model parameters to experimental setup}

While our work is a theoretical model study, it is important to compare our results to the experiments and search for possible relevance. By matching the model parameters to the experimental setup, we can find the effective regime of our results and propose the phenomena to search for. In this section, we discuss the relevant comparisons with the pump-probe experiments of kagome metals $A$V$_3$Sb$_5$ with $A=$ K, Rb, and Cs.

\subsection{Unit conversion}

To compare the theoretical model to the experiments, a crucial procedure is the unit conversion. We begin with the units of the energy and the time. For the energy unit, it is natural to assume
\begin{equation}
1 \text{ energy unit}\equiv1\text{ eV}\approx1.602\times10^{-19} \text{ J}.
\end{equation}
This assumption places the band structure scale in the realistic regime of the kagome metals $A$V$_3$Sb$_5$. By considering the photon-energy relation
\begin{equation}
1 \text{ energy unit}=\hbar\times(1 \text{ frequency unit}),
\end{equation}
we know that the numerical values of the frequency $\omega$ (more precisely, the angular velocity) exactly represents its corresponding energy in the unit of eV. We can further convert the frequency unit to THz, which is useful for the identification of collective-mode oscillations
\begin{equation}
1 \text{ frequency unit}=\frac{1 \text{ energy unit}}{\hbar}\equiv\frac{1\text{ eV}}{\hbar}\approx\frac{1.602\times10^{-19} \text{ J}}{1.055\times10^{-34} \text{ J}\cdot\text{s}}\approx1518.483\text{ ps}^{-1}=\frac{1518.483}{2\pi}\text{ THz}\approx241.7\text{ THz}.
\end{equation}
The frequency relation also implies a conversion of the time unit
\begin{equation}
1 \text{ time unit}
=\frac{1}{1 \text{ frequency unit}}
\approx\frac{1}{1518.483\text{ ps}^{-1}}
\approx6.586\times10^{-4} \text{ ps}
\approx0.659 \text{ fs},
\end{equation}
which is at the femtosecond (fs) scale. The time evolution up to $t=500$ in our analysis then represents a period of $0.329$ picosecond (ps). On the other hand, it is also important to specify the length unit. In our model, we assume the nearest-neighbor site distance $a=1$. Based on the practical structure of CsV$_3$Sb$_5$ \cite{tsirlin22sp}, we set
\begin{equation}
1 \text{ length unit}\approx2.75\times10^{-10} \text{ m}.
\end{equation}

\subsection{Setup of pump pulse}

The pump-probe experiments involve the application of laser pump pulses. To ensure that our analysis introduces reasonable pump pulses to the model, we need to check whether our parameters convert to reasonable values in the experiments. A pump pulse can be defined by a time-dependent gauge field
\begin{equation}
\label{suppeq:pumppulse}
\mbf A(t)=A_\text{c}e^{-(t-t_\text{c})^2/2\sigma_\text{t}^2}\mbf e_\text{p}(t,\omega_\text{c}).
\end{equation}
The center time of the peak is set as $t_\text{c}=25$, and a chosen characteristic peak width $\sigma_\text{t}=3$ leads to a pulse duration $\approx[t_c-25,t_c+25]$ with accessible amplitude $\exp[-(t-t_\text{c})^2/2\sigma_\text{t}^2]>10^{-15}$. Note that this duration corresponds to $50\times0.659\text{ fs}\approx32.9\text{ fs}$, which is in the relevant range of the experiments. The tunable parameters are the center amplitude $A_\text{c}$, the center frequency $\omega_\text{c}$, and the oscillating polarization vector $\mbf e_\text{p}(t,\omega_\text{c})$. We have known the identical conversion from the frequency unit to eV. For example, a pump pulse with $\omega_\text{c}=2$ carries the energy $2$ eV. by choosing $\omega_\text{c}\in[1,3]$, the pump pulses fall in the accessible ranges from the $1.55$ eV Ti:sapphire laser sources \cite{giannetti16aip}. Meanwhile, the fluence deserves a careful calculation.

Before calculating the fluence of the specific pump pulses, we begin with the unit conversion for the general lasers. The lasers can be described by a time-dependent electric field $\mbf E(t)$. Since the unit of the electric field is $\text{V}/\text{m}=\text{J}/(\text{C}\cdot\text{m})$, the unit strength of the electric field in our model is
\begin{equation}
1 \text{ electric-field unit}
\equiv\frac{1 \text{ energy unit}}{1 \text{ charge unit}\times1 \text{ length unit}}
=\frac{1.602\times10^{-19} \text{ J}}{1.602\times10^{-19} \text{ C}\times2.75\times10^{-10} \text{ m}}
\approx3.636\times10^9 \text{ J}/(\text{C}\cdot\text{m}).
\end{equation}
The fluence is the total energy delivered by a laser in the laser-shining period
\begin{equation}
F=\int dt\frac{1}{2}c\epsilon_0|\mbf E(t)|^2.
\end{equation}
Here the integrand represents the optical intensity $I(t)$, which is the optical energy flux carried by the laser. For simplicity, we have assumed a vacuum environment, which sets the permittivity in its vacuum value $\epsilon_0$ and neglects any frequency-dependent refraction effects in the intensity $I(t)$. We want to derive the unit conversion of the fluence to $\text{mJ}/\text{cm}^2$. The resulting formula can be obtained as
\begin{equation}
\begin{aligned}
1\text{ fluence unit}
&\equiv\frac{1}{2}c\epsilon_0\times(1\text{ electric-field unit})^2\times(10^3\text{ mJ}/\text{J})\times(10^{-2}\text{ m}/\text{cm})^2\times(1\text{ time unit})\\
&\approx\frac{1}{2}\times(3\times10^8\text{ m}/\text{s})\times[8.854\times10^{-12}\text{ C}^2/(\text{J}\cdot\text{m})]\times[3.636\times10^9 \text{ J}/(\text{C}\cdot\text{m})]^2\times(10^3\text{ mJ}/\text{J})\\&\quad\times(10^{-2}\text{ m}/\text{cm})^2\times(6.586\times10^{-16}\text{ s})\\
&\approx1.156\text{ mJ}/\text{cm}^2.
\end{aligned}
\end{equation}
Therefore, the fluence in our analysis can be converted as
\begin{equation}
F=\left[\int dt|\mbf E(t)|^2\right]\times(1\text{ fluence unit})
=\left[\int dt|\mbf E(t)|^2\right]\times(1.156\text{ mJ}/\text{cm}^2).
\end{equation}

\begin{figure}[t]
\centering
\includegraphics[scale=0.55]{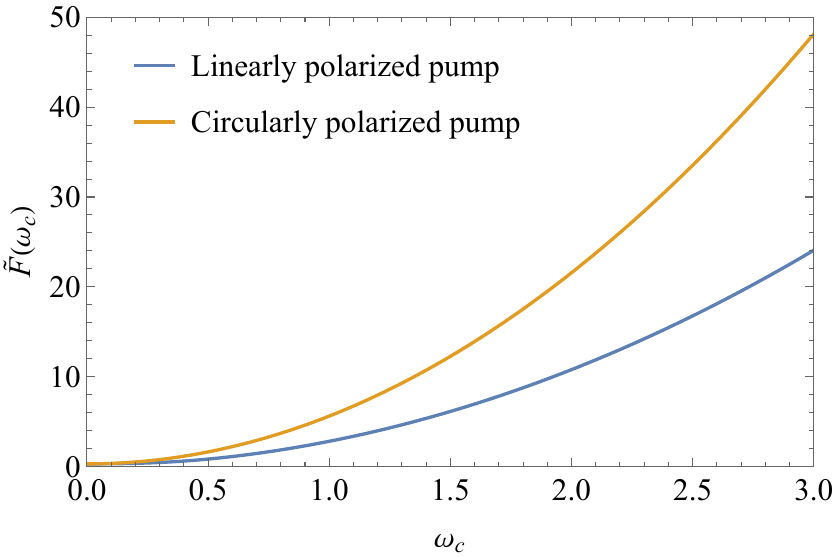}
\caption{\label{suppfig:fluence} The integral $\tilde F(\omega_\text{c})$ in the fluence $F(A_\text{c},\omega_\text{c})$.}
\end{figure}

We can now compute the fluences of the pump pulses in our analysis. An electric field can be derived from the gauge field (\ref{suppeq:pumppulse})
\begin{equation}
\mbf E(t)=-\frac{\partial\mbf A(t)}{\partial t}
=-A_\text{c}e^{-(t-t_\text{c})^2/2\sigma_\text{t}^2}\left[-\frac{t-t_\text{c}}{\sigma_\text{t}^2}\mbf e_\text{p}(t,\omega_\text{c})+\frac{\partial}{\partial t}\mbf e_\text{p}(t,\omega_\text{c})\right],
\end{equation}
which leads to a fluence
\begin{equation}
F(A_\text{c},\omega_\text{c})=A_\text{c}^2\tilde F(\omega_\text{c})\times(1.156\text{ mJ}/\text{cm}^2),\quad
\tilde F(\omega_\text{c})=\int dte^{-(t-t_\text{c})^2/\sigma_\text{t}^2}\left|-\frac{t-t_\text{c}}{\sigma_\text{t}^2}\mbf e_\text{p}(t,\omega_\text{c})+\frac{\partial}{\partial t}\mbf e_\text{p}(t,\omega_\text{c})\right|^2.
\end{equation}
The integral $\tilde F(\omega_\text{c})$ depends on the polarization of the pump pulse (Fig.~\ref{suppfig:fluence}). For the linearly polarized pump, the oscillating polarization vector $\mbf e_\text{p}(t,\omega_\text{c})=\cos[\omega_\text{c}(t-t_\text{c})]\mbf e_{\text{p},0}$ points along the direction of a unit vector $\mbf e_{\text{p},0}$. With the oscillating part
\begin{equation}
-\frac{t-t_\text{c}}{\sigma_\text{t}^2}\mbf e_\text{p}(t,\omega_\text{c})+\frac{\partial}{\partial t}\mbf e_\text{p}(t,\omega_\text{c})
=\left\{-\frac{t-t_\text{c}}{\sigma_\text{t}^2}\cos[\omega_\text{c}(t-t_\text{c})]-\omega_\text{c}\sin[\omega_\text{c}(t-t_\text{c})]\right\}\mbf e_{\text{p},0},
\end{equation}
the integral $\tilde F(\omega_\text{c})$ can be computed straightforwardly. For the frequency range $\omega_\text{c}\in[1,3]$ in our analysis, the integral falls in the range $\tilde F(\omega_\text{c})\in[2.801,24.076]$. With the center amplitude $A_\text{c}\in[0.02,0.1]$, the fluence falls in the relevant range of the experiments $F(A_\text{c},\omega_\text{c})\in[0.0013,0.278]$ with the unit $\text{mJ}/\text{cm}^2$. Similarly, the circularly polarized pump manifests the time-dependent polarization vector $\mbf e_\text{p}(t,\omega_\text{c})=(\cos[\omega_\text{c}(t-t_\text{c})],\sin[\omega_\text{c}(t-t_\text{c})],0)$. This structure gives the oscillating part
\begin{equation}
\begin{aligned}
&-\frac{t-t_\text{c}}{\sigma_\text{t}^2}\mbf e_\text{p}(t,\omega_\text{c})+\frac{\partial}{\partial t}\mbf e_\text{p}(t,\omega_\text{c})\\
&=\left(-\frac{t-t_\text{c}}{\sigma_\text{t}^2}\cos[\omega_\text{c}(t-t_\text{c})]-\omega_\text{c}\sin[\omega_\text{c}(t-t_\text{c})],-\frac{t-t_\text{c}}{\sigma_\text{t}^2}\sin[\omega_\text{c}(t-t_\text{c})]+\omega_\text{c}\cos[\omega_\text{c}(t-t_\text{c})],0\right),
\end{aligned}
\end{equation}
and the corresponding integral can be calculated analogously $\tilde F(\omega_\text{c})\in[5.613,48.152]$. The corresponding fluence again falls in the relevant range of the experiments $F(A_\text{c},\omega_\text{c})\in[0.0026,0.557]$ with the unit $\text{mJ}/\text{cm}^2$.

\section{Time-dependent Hartree-Fock theory}

With the Hartree-Fock theory, we can study various properties of the ground states in the interacting-fermion systems. Meanwhile, it is also interesting, and sometimes important, to understand how the interaction-driven states evolve under the time-varying conditions. In this section, we discuss how the dynamics of an interacting system can be investigated in the time-dependent Hartree-Fock theory.

\subsection{General formalism}

We again consider the general model (\ref{suppeq:model0}) of the interacting-fermion systems, but now in a time-dependent form
\begin{equation}
H(t)=\sum_{ab}\mathcal H_{0,ab}(t)c_a^\dagger c_b+\frac{1}{2}\sum_{abcd}U_{acdb}(t)c_a^\dagger c_c^\dagger c_dc_b.
\end{equation}
Different from the static systems, here we allow the time dependence in the noninteracting Hamiltonian $\mathcal H_0(t)$ and the interaction $U(t)$. Given a $N$-fermion state $\ket{\Psi(t)}$, its time evolution is governed by the Schr\"odinger equation
\begin{equation}
i\frac{d}{dt}\ket{\Psi(t)}=H(t)\ket{\Psi(t)}.
\end{equation}
To formulate the time-dependent Hartree-Fock theory in terms of the density matrix
\begin{equation}
P_{ab}(t)=\braket{\Psi(t)}{c_b^\dagger c_a}{\Psi(t)},
\end{equation}
we need to derive the equation that captures its time evolution
\begin{equation}
\begin{aligned}
i\frac{d}{dt}P_{ab}
&=i\frac{d}{dt}\braket{\Psi}{c_b^\dagger c_a}{\Psi}
=\left(i\frac{d}{dt}\bra{\Psi}\right)c_b^\dagger c_a\ket{\Psi}+\bra{\Psi}c_b^\dagger c_a\left(i\frac{d}{dt}\ket{\Psi}\right)\\
&=(-\bra{\Psi}H)c_b^\dagger c_a\ket{\Psi}+\bra{\Psi}c_b^\dagger c_a(H\ket{\Psi})
=\epvl{-Hc_b^\dagger c_a+c_b^\dagger c_aH}\\
&=\epvl{[c_b^\dagger c_a,H]}.
\end{aligned}
\end{equation}
Adopting the explicit form of the Hamiltonian $H$, we get
\begin{equation}
\label{suppeq:timepcomm}
\begin{aligned}
i\frac{d}{dt}P_{ab}
=\Epvl{\left[c_b^\dagger c_a,\sum_{cd}\mathcal H_{0,cd}c_c^\dagger c_d+\frac{1}{2}\sum_{cdef}U_{cefd}c_c^\dagger c_e^\dagger c_fc_d\right]}
=\sum_{cd}\mathcal H_{0,cd}\epvl{[c_b^\dagger c_a,c_c^\dagger c_d]}+\frac{1}{2}\sum_{cdef}U_{cefd}\epvl{[c_b^\dagger c_a,c_c^\dagger c_e^\dagger c_fc_d]}.
\end{aligned}
\end{equation}
Importantly, we assume that the state $\ket{\Psi(t)}$ remains in the mean-field regime as in the static formalism. Under this condition, we are able to derive the time-evolution equation of the density matrix.

The time-evolution equation involves various commutators between the fermionic operators. To reduce this equation, we exploit the useful reduction relations of the commutator $[AB,C]=A[B,C]+[A,C]B=A\{B,C\}-\{A,C\}B$ and the anticommutation relations of the fermionic operators $\{c_a^\dagger,c_b\}=\delta_{ab}$ and $\{c_a,c_b\}=\{c_a^\dagger,c_b^\dagger\}=0$. We first derive the contribution from the noninteracting Hamiltonian. By calculating the commutator between the particle-hole bilinears
\begin{equation}
\begin{aligned}
[c_a^\dagger c_b,c_c^\dagger c_d]
&=c_c^\dagger[c_a^\dagger c_b,c_d]+[c_a^\dagger c_b,c_c^\dagger]c_d\\
&=c_c^\dagger(c_a^\dagger\{c_b,c_d\}-\{c_a^\dagger,c_d\}c_b)+(c_a^\dagger\{c_b,c_c^\dagger\}-\{c_a^\dagger,c_c^\dagger\}c_b)c_d\\
&=-\delta_{ad}c_c^\dagger c_b+\delta_{bc}c_a^\dagger c_d,
\end{aligned}
\end{equation}
we obtain the result
\begin{equation}
\label{suppeq:timenint}
\begin{aligned}
\sum_{cd}\mathcal H_{0,cd}\epvl{[c_b^\dagger c_a,c_c^\dagger c_d]}
&=\sum_{cd}\mathcal H_{0,cd}\epvl{-\delta_{bd}c_c^\dagger c_a+\delta_{ac}c_b^\dagger c_d}
=-\sum_c\mathcal H_{0,cb}\epvl{c_c^\dagger c_a}+\sum_d\mathcal H_{0,ad}\epvl{c_b^\dagger c_d}\\
&=\sum_c(-\mathcal H_{0,cb}\epvl{c_c^\dagger c_a}+\mathcal H_{0,ac}\epvl{c_b^\dagger c_c})
=\sum_c(-\mathcal H_{0,cb}P_{ac}+\mathcal H_{0,ac}P_{cb})\\
&=[\mathcal H_0,P]_{ab}.
\end{aligned}
\end{equation}
We proceed to calculate the contribution from the interaction. With the reduction relations of the commutator, the interaction contribution can be reduced
\begin{equation}
\begin{aligned}
\frac{1}{2}\sum_{cdef}U_{cefd}\epvl{[c_b^\dagger c_a,c_c^\dagger c_e^\dagger c_fc_d]}
&=\frac{1}{2}\sum_{cdef}U_{cefd}\epvl{c_c^\dagger c_e^\dagger[c_b^\dagger c_a,c_fc_d]+[c_b^\dagger c_a,c_c^\dagger c_e^\dagger]c_fc_d}\\
&=\frac{1}{2}\sum_{cdef}U_{cefd}\epvl{c_c^\dagger c_e^\dagger[c_b^\dagger,c_fc_d]c_a+c_b^\dagger[c_a,c_c^\dagger c_e^\dagger]c_fc_d}.
\end{aligned}
\end{equation}
In the second line, we have used the fact that $[c_a,c_bc_c]=[c_a^\dagger,c_b^\dagger c_c^\dagger]=0$. A further reduction leads to
\begin{equation}
\begin{aligned}
\frac{1}{2}\sum_{cdef}U_{cefd}\epvl{[c_b^\dagger c_a,c_c^\dagger c_e^\dagger c_fc_d]}
&=\frac{1}{2}\sum_{cdef}U_{cefd}\epvl{c_c^\dagger c_e^\dagger(-c_f\{c_b^\dagger,c_d\}+\{c_b^\dagger,c_f\}c_d)c_a+c_b^\dagger(-c_c^\dagger\{c_a,c_e^\dagger\}+\{c_a,c_c^\dagger\}c_e^\dagger)c_fc_d}\\
&=\frac{1}{2}\sum_{cdef}U_{cefd}\epvl{c_c^\dagger c_e^\dagger(-c_f\delta_{bd}+\delta_{bf}c_d)c_a+c_b^\dagger(-c_c^\dagger\delta_{ae}+\delta_{ac}c_e^\dagger)c_fc_d}\\
&=-\frac{1}{2}\sum_{cef}U_{cefb}\epvl{c_c^\dagger c_e^\dagger c_fc_a}+\frac{1}{2}\sum_{cde}U_{cebd}\epvl{c_c^\dagger c_e^\dagger c_dc_a}\\&\quad-\frac{1}{2}\sum_{cdf}U_{cafd}\epvl{c_b^\dagger c_c^\dagger c_fc_d}+\frac{1}{2}\sum_{def}U_{aefd}\epvl{c_b^\dagger c_e^\dagger c_fc_d}\\
&=\frac{1}{2}\sum_{cde}(-U_{cdeb}\epvl{c_c^\dagger c_d^\dagger c_ec_a}+U_{dcbe}\epvl{c_d^\dagger c_c^\dagger c_ec_a}-U_{caed}\epvl{c_b^\dagger c_c^\dagger c_ec_d}+U_{acde}\epvl{c_b^\dagger c_c^\dagger c_dc_e}).
\end{aligned}
\end{equation}
Since the interaction obeys the permutation rule $U_{acdb}=U_{cabd}$, the equation becomes
\begin{equation}
\begin{aligned}
\frac{1}{2}\sum_{cdef}U_{cefd}\epvl{[c_b^\dagger c_a,c_c^\dagger c_e^\dagger c_fc_d]}
&=\frac{1}{2}\sum_{cde}[U_{cdeb}(-\epvl{c_c^\dagger c_d^\dagger c_ec_a}+\epvl{c_d^\dagger c_c^\dagger c_ec_a})+U_{acde}(-\epvl{c_b^\dagger c_c^\dagger c_ec_d}+\epvl{c_b^\dagger c_c^\dagger c_dc_e})]\\
&=\sum_{cde}(-U_{cdeb}\epvl{c_c^\dagger c_d^\dagger c_ec_a}+U_{acde}\epvl{c_b^\dagger c_c^\dagger c_dc_e}).
\end{aligned}
\end{equation}
From the Wick's theorem (\ref{suppeq:wick}), we further get
\begin{equation}
\begin{aligned}
\frac{1}{2}\sum_{cefd}U_{cefd}\epvl{[c_b^\dagger c_a,c_c^\dagger c_e^\dagger c_fc_d]}
&=\sum_{cde}[-U_{cdeb}(P_{ac}P_{ed}-P_{ec}P_{ad})+U_{acde}(P_{eb}P_{dc}-P_{db}P_{ec})]\\
&=\sum_{cde}[-P_{ac}(U_{cdeb}-U_{dceb})P_{ed}+(U_{adec}-U_{adce})P_{ed}P_{cb}]\\
&=\sum_{cde}\{-P_{ac}[(U_{cdeb}-U_{cdbe})P_{ed}]+[(U_{adec}-U_{adce})P_{ed}]P_{cb}\}.
\end{aligned}
\end{equation}
Remarkably, this equation can be rewritten in terms of the Hartree-Fock potential $\mathcal V_\text{HF}[P(t)]$ (\ref{suppeq:hfham})
\begin{equation}
\label{suppeq:timeint}
\begin{aligned}
\frac{1}{2}\sum_{cdef}U_{cefd}\epvl{[c_b^\dagger c_a,c_c^\dagger c_e^\dagger c_fc_d]}
&=\sum_c[-P_{ac}\mathcal V_\text{HF}[P]_{cb}+\mathcal V_\text{HF}[P]_{ac}P_{cb}]\\
&=[\mathcal V_\text{HF}[P],P]_{ab}.
\end{aligned}
\end{equation}
Combining the contributions from the noninteracting Hamiltonian (\ref{suppeq:timenint}) and the interaction (\ref{suppeq:timeint}), the original equation (\ref{suppeq:timepcomm}) acquires a simple form
\begin{equation}
i\frac{d}{dt}P_{ab}
=[\mathcal H_0,P]_{ab}+[\mathcal V_\text{HF}[P],P]_{ab}
=[\mathcal H_\text{HF}[P],P]_{ab}.
\end{equation}
Therefore, the time evolution of the density matrix is described by the equation
\begin{equation}
\label{suppeq:timepeq}
i\frac{d}{dt}P(t)=[\mathcal H_\text{HF}[P(t)](t),P(t)].
\end{equation}

Since the Hartree-Fock theory is effectively noninteracting at each instant, the time-evolution equation (\ref{suppeq:timepeq}) takes the simple form of the von Neumann equation. With this simplicity, the nonequilibrium dynamics can be studied by the time-dependent Hartree-Fock theory in an efficient manner. Meanwhile, the interaction links the Hartree-Fock Hamiltonian $\mathcal H_\text{HF}[P(t)](t)$ to the time evolution of the density matrix $P(t)$. This interaction effect generally drives the time evolution nonlinear, and complicated phenomena may occur along the way. On the other hand, one should remember that the mean-field theory does not include all of the many-body relaxation processes. At strong nonequilibrium, the ignored processes may lead to significant deviation from the time-dependent Hartree-Fock theory.

\subsection{Important properties}

Before diving into the practical implementation of the time-dependent Hartree-Fock theory, we discuss some properties which turn out to be important in the numerical computations.

\subsubsection{Projector condition}

In the time-dependent Hartree-Fock theory, an important property is the preservation of projector condition $P^2=P$ under the time evolution (\ref{suppeq:timepeq}). Assume that the density matrix $P(t)$ obeys the projector condition $P^2(t)-P(t)=0$ at the time $t$. Examining the time derivative of the projector difference $P^2-P$
\begin{equation}
\begin{aligned}
\frac{d}{dt}(P^2-P)
&=P\frac{dP}{dt}+\frac{dP}{dt}P-\frac{dP}{dt}
=P\left(\frac{1}{i}[\mathcal H_{HF}[P],P]\right)+\left(\frac{1}{i}[\mathcal H_{HF}[P],P]\right)P-\frac{dP}{dt}\\
&=\frac{1}{i}[\mathcal H_{HF}[P],P^2]-\frac{dP}{dt}
=\frac{1}{i}[\mathcal H_{HF}[P],P]-\frac{dP}{dt}\\
&=0,
\end{aligned}
\end{equation}
we find that it remains zero as the time changes. Therefore, if the initial density matrix $P(t=0)$ obeys the projector condition, the condition will always be valid under the time evolution.

\subsubsection{Energy change}

Another important property is the energy change
\begin{equation}
\begin{aligned}
\frac{d}{dt}E[P]
&=\frac{d}{dt}\left[\frac{1}{2}\Tr(P(\mathcal H_0+\mathcal H_\text{HF}[P]))\right]
=\frac{d}{dt}\Tr\left(P\left(\mathcal H_0+\frac{1}{2}\mathcal V_\text{HF}[P]\right)\right)\\
&=\Tr\left(\frac{dP}{dt}\left(\mathcal H_0+\frac{1}{2}\mathcal V_\text{HF}[P]\right)\right)+\Tr\left(P\left(\frac{d\mathcal H_0}{dt}+\frac{1}{2}\frac{d\mathcal V_\text{HF}[P]}{dt}\right)\right)\\
&=\Tr\left(\frac{dP}{dt}\left(\mathcal H_0+\frac{1}{2}\mathcal V_\text{HF}[P]\right)\right)+\Tr\left(P\left(\frac{d\mathcal H_0}{dt}+\frac{1}{2}\frac{d\mathcal V_\text{HF}}{dt}[P]+\frac{1}{2}\mathcal V_\text{HF}\left[\frac{dP}{dt}\right]\right)\right).
\end{aligned}
\end{equation}
The first two terms in the second trace result from the time dependence of the noninteracting Hamiltonian $\mathcal H_0(t)$ and the interaction $U(t)$, respectively. Adopting the relation $\Tr(P_1\mathcal V_\text{HF}[P_2])=\Tr(P_2\mathcal V_\text{HF}[P_1])$, our calculation proceeds as
\begin{equation}
\begin{aligned}
\frac{d}{dt}E[P]
&=\Tr\left(\frac{dP}{dt}\left(\mathcal H_0+\frac{1}{2}\mathcal V_\text{HF}[P]\right)\right)+\frac{1}{2}\Tr\left(\frac{dP}{dt}\mathcal V_\text{HF}[P]\right)+\Tr\left(P\left(\frac{d\mathcal H_0}{dt}+\frac{1}{2}\frac{d\mathcal V_\text{HF}}{dt}[P]\right)\right)\\
&=\Tr\left(\frac{dP}{dt}(\mathcal H_0+\mathcal V_\text{HF}[P])\right)+\Tr\left(P\left(\frac{d\mathcal H_0}{dt}+\frac{1}{2}\frac{d\mathcal V_\text{HF}}{dt}[P]\right)\right)\\
&=\Tr\left(\frac{dP}{dt}\mathcal H_\text{HF}[P]\right)+\Tr\left(P\left(\frac{d\mathcal H_0}{dt}+\frac{1}{2}\frac{d\mathcal V_\text{HF}}{dt}[P]\right)\right).
\end{aligned}
\end{equation}
Since the insertion of the time-evolution equation (\ref{suppeq:timepeq}) eliminates the first term
\begin{equation}
\Tr\left(\frac{dP}{dt}\mathcal H_\text{HF}[P]\right)=\Tr\left(\frac{1}{i}[\mathcal H_\text{HF}[P],P]\mathcal H_\text{HF}[P]\right)=0,
\end{equation}
we arrive at the final result
\begin{equation}
\frac{d}{dt}E[P]
=\Tr\left(P\left(\frac{d\mathcal H_0}{dt}+\frac{1}{2}\frac{d\mathcal V_\text{HF}}{dt}[P]\right)\right).
\end{equation}
Importantly, the energy is conserved if the Hamiltonian is time-independent.

As we will see later, the projector condition and the energy conservation will play important roles in the validity of the numerical implementation.

\subsubsection{Current orders}
\label{suppsssec:current}

As mentioned previously in Sec.~\ref{suppsssec:hfcharge}, the current flowing between different states can be an important symmetry-breaking order. Moreover, the time evolution is achieved through these currents, so it is important to understand how to quantify them. Define the current $j_{ab}$ as the flow from a state $a$ to another state $b$. We consider the conservation law at a specific state $b$, where the density $P_{bb}$ varies by receiving the currents $j_{ab}$ from the other states $a$
\begin{equation}
\begin{aligned}
\sum_{a}j_{ab}
&=\frac{d}{dt}P_{bb}
=\frac{1}{i}[\mathcal H_\text{HF}[P],P]_{bb}
=-i(\mathcal H_\text{HF}[P]P-P\mathcal H_\text{HF}[P])_{bb}
=-i\sum_c(\mathcal H_\text{HF}[P]_{bc}P_{cb}-P_{bc}\mathcal H_\text{HF}[P]_{cb})\\
&=-i\sum_c\left[\left(\mathcal H_{0,{bc}}+\sum_{de}(U_{bdec}-U_{bdce})P_{ed}\right)P_{cb}-P_{bc}\left(\mathcal H_{0,{cb}}+\sum_{de}(U_{cdeb}-U_{cdbe})P_{ed}\right)\right]\\
&=-i\sum_c(\mathcal H_{0,{bc}}P_{cb}-P_{bc}\mathcal H_{0,{cb}})-i\sum_{cde}[(U_{bdec}-U_{bdce})P_{ed}P_{cb}-(U_{cdeb}-U_{cdbe})P_{ed}P_{bc}]\\
&=\sum_a\left\{2\text{Im}(\mathcal H_{0,{ba}}P_{ab})-i\sum_{cd}[(U_{bcda}-U_{bcad})P_{dc}P_{ab}-(U_{acdb}-U_{acbd})P_{dc}P_{ba}]\right\}.
\end{aligned}
\end{equation}
The current can thus be written in terms of the density matrix
\begin{equation}
j_{ab}=2\text{Im}(\mathcal H_{0,{ba}}P_{ab})-i\sum_{cd}[(U_{bcda}-U_{bcad})P_{dc}P_{ab}-(U_{acdb}-U_{acbd})P_{dc}P_{ba}].
\end{equation}
The first term directly tells us the contribution from the noninteracting Hamiltonian. Meanwhile, the second term shows the possible correction from the interactions. Notably, if the interactions take the density-density form $U_{acdb}=U_{acca}\delta_{ab}\delta_{cd}$ with $U_{acca}=U_{caac}$, its contribution vanishes
\begin{equation}
\begin{aligned}
&-i\sum_{cd}[(U_{bcda}-U_{bcad})P_{dc}P_{ab}-(U_{acdb}-U_{acbd})P_{dc}P_{ba}]\\
&=-i\left[\delta_{ab}\sum_c(U_{bccb}P_{cc}P_{bb}-U_{bccb}P_{cc}P_{bb})-U_{baab}P_{ba}P_{ab}+U_{abba}P_{ab}P_{ba}\right]\\
&=0
\end{aligned}
\end{equation}
and leaves only the noninteracting-Hamiltonian contribution in the current $j_{ab}$. This result implies, for example, the relation (\ref{suppeq:current}) between the charge-current density and the imaginary charge-bond order in the Hubbard model.

\subsection{Numerical scheme}

Having presented the general formalism of the time-dependent Hartree-Fock theory, we now discuss its practical implementation as a numerical method.

\subsubsection{Defects in naive discrete time evolution}

In the time-dependent Hartree-Fock theory, the time-evolution equation (\ref{suppeq:timepeq}) is a differential equation of a continuous time $t$. However, in the numerical implementation, we need to introduce a small discrete time step $\Delta t$. As the time goes forward $t\rightarrow t+\Delta t$, one may expect the time evolution to be captured by the difference equation
\begin{equation}
\label{suppeq:wrongdistimeeq}
\Delta P(t)=P(t+\Delta t)-P(t)=\Delta t\frac{dP(t)}{dt}=-i\Delta t[\mathcal H_\text{HF}[P(t)](t),P(t)].
\end{equation}
This discretized formalism indeed resumes the original one (\ref{suppeq:timepeq}) in the continuous time limit $\Delta t\rightarrow0$. However, a serious defect occurs that the projector condition $P^2=P$ fails for any discrete time step $\Delta t>0$. To observe this defect, we apply an evolution $t\rightarrow t+\Delta t$ to a density matrix $P(t)$ which obeys the projector condition $P(t)^2-P(t)=0$ at the time $t$. The projector difference $P^2-P$ evolves as
\begin{equation}
\begin{aligned}
P(t+\Delta t)^2-P(t+\Delta t)
&=[P(t)+\Delta P(t)]^2-[P(t)+\Delta P(t)]\\
&=[P(t)^2+P(t)\Delta P(t)+\Delta P(t)P(t)+\Delta P(t)^2]-[P(t)+\Delta P(t)]\\
&=[P(t)^2-P(t)]+[-\Delta P(t)+P(t)\Delta P(t)+\Delta P(t)P(t)+\Delta P(t)^2]\\
&=\Delta P(t)^2-\Delta P(t)+P(t)\Delta P(t)+\Delta P(t)P(t).
\end{aligned}
\end{equation}
Adopting the difference equation (\ref{suppeq:wrongdistimeeq}), we get
\begin{equation}
\begin{aligned}
P(t+\Delta t)^2-P(t+\Delta t)
&=\Delta P(t)^2-\Delta P(t)-i\Delta t(P(t)[\mathcal H_\text{HF}[P(t)](t),P(t)]+[\mathcal H_\text{HF}[P(t)](t),P(t)]P(t))\\
&=\Delta P(t)^2-\Delta P(t)-i\Delta t[\mathcal H_\text{HF}[P(t)](t),P(t)^2]\\
&=\Delta P(t)^2-\Delta P(t)-i\Delta t[\mathcal H_\text{HF}[P(t)],P(t)]\\
&=\Delta P(t)^2.
\end{aligned}
\end{equation}
This result clearly breaks the projector condition at the order of $\Delta P(t)^2$.

The energy conservation is also broken under the difference equation (\ref{suppeq:wrongdistimeeq}). Assuming that the Hamiltonian is time-independent, we consider the energy difference at two adjacent time steps
\begin{equation}
\begin{aligned}
\Delta E
&=E[P(t+\Delta t)]-E[P(t)]
=\frac{1}{2}\Tr(P(t+\Delta t)(\mathcal H_0+\mathcal H_\text{HF}[P(t+\Delta t)]))-\frac{1}{2}\Tr(P(t)(\mathcal H_0+\mathcal H_\text{HF}[P(t)]))\\
&=\Tr\left([P(t)+\Delta P(t)]\left(\mathcal H_0+\frac{1}{2}\mathcal V_\text{HF}[P(t)+\Delta P(t)]\right)\right)-\Tr\left(P(t)\left(\mathcal H_0+\frac{1}{2}\mathcal V_\text{HF}[P(t)]\right)\right)\\
&=\Tr\left(\Delta P(t)\left(\mathcal H_0+\frac{1}{2}\mathcal V_\text{HF}[P(t)]\right)\right)+\frac{1}{2}\Tr(P(t)\mathcal V_\text{HF}[\Delta P(t)])+\frac{1}{2}\Tr(\Delta P(t)\mathcal V_\text{HF}[\Delta P(t)])\\
&=\Tr\left(\Delta P(t)\left(\mathcal H_0+\frac{1}{2}\mathcal V_\text{HF}[P(t)]\right)\right)+\frac{1}{2}\Tr(\Delta P(t)\mathcal V_\text{HF}[P(t)])+\frac{1}{2}\Tr(\Delta P(t)\mathcal V_\text{HF}[\Delta P(t)])\\
&=\Tr(\Delta P(t)\mathcal H_\text{HF}[P(t)])+\frac{1}{2}\Tr(\Delta P(t)\mathcal V_\text{HF}[\Delta P(t)])\\
&=-i\Delta t\Tr([\mathcal H_\text{HF}[P(t)],P(t)]\mathcal H_\text{HF}[P(t)])+\frac{1}{2}\Tr(\Delta P(t)\mathcal V_\text{HF}[\Delta P(t)])\\
&=\frac{1}{2}\Tr(\Delta P(t)\mathcal V_\text{HF}[\Delta P(t)]).
\end{aligned}
\end{equation}
We observe a breakdown of the energy conservation, with an error again at the order of $\Delta P(t)^2$.

The difference of the density matrix is proportional to the discrete time step $\Delta P(t)\sim\Delta t$. As such, the errors in both the projector condition and the energy conservation are quadratic in the discrete time step $O(\Delta t^2)$. If we stay with the difference equation (\ref{suppeq:wrongdistimeeq}), the error reduction requires an extremely small time step and thus very long computational time. It is necessary to find an alternative approach to resolve these defects.

\subsubsection{Runge-Kutta method}

A powerful resolution to the discrete time errors is the famous Runge-Kutta method \Scite{atkinson91book,press07book,butcher16book}. Here we adopt its fourth order version, known as the RK4. For conceptual clarity, we first describe it in terms of a general ordinary differential equation of a function $y(t)$
\begin{equation}
\frac{dy}{dt}=f(t,y)
\end{equation}
with an initial condition $y(t=0)=y_0$. For each time step $t_m\rightarrow t_{m+1}=t_m+\Delta t$, the RK4 method moves $y_m$ to $y_{m+1}$ as follows:
\begin{enumerate}
\item Compute $k_1=f(t_m,y_m)$.
\item Compute $k_2=f(t_m+\Delta t/2,y_m+[\Delta t/2]k_1)$.
\item Compute $k_3=f(t_m+\Delta t/2,y_m+[\Delta t/2]k_2)$.
\item Compute $k_4=f(t_m+\Delta t,y_m+\Delta tk_3)$.
\item Compute $y_{m+1}=y_m+(\Delta t/6)(k_1+2k_2+2k_3+k_4)$.
\end{enumerate}
We can see that the RK4 method averages the slopes at the tentative middle and end points, thereby reducing the error from the discrete time step. Remarkably, it has been proven that the design of the RK4 method can reduce the error to an order $O(\Delta t^5)$ in each time step. This reduction significantly improves the accuracy of the discrete time evolution at reasonable time steps.

Now we adopt the RK4 method in the time-dependent Hartree-Fock theory. We identify the density matrix $P_m$ at time $t_m$ with the general function $y_m$, and use the time-evolution commutator $-i[\mathcal H_\text{HF}[P](t),P]$ as the derivative $f(t,y)$. As we discuss later, the RK4 method yields amazingly small errors in both the projector condition and the energy conservation.

\subsubsection{Algorithm}

The time-dependent Hartree-Fock theory is again implemented as an iterative algorithm. The time evolution begins with the initial iteration $m=0$, where an initial density matrix $P_m=P(t_m=0)$ is assigned. To ensure the validity of the results, a small enough time step is needed. If the initial state is the ground state, a quick reference can be acquired by examining its time invariance under the time-independent Hamiltonian. Usually, a time step $\Delta t$ at the order $O(10^{-2})$ is good enough. Each iteration aims at evolving the density matrix $P_m$ at the time $t_m$ to a new one $P_{m+1}$ at the next time $t_{m+1}=t_m+\Delta t$:
\begin{enumerate}
\item Get the time-dependent noninteracting Hamiltonian $\mathcal H_{0,m}=\mathcal H_0(t_m)$ and the interaction $U_m=U(t_m)$.
\item Compose the Hartree-Fock Hamiltonian $\mathcal H_{\text{HF},m}=\mathcal H_\text{HF}[P_m](t_m)$.
\item Compute the desired physical observables, including the energy.
\item Perform the discrete time evolution by the RK4 method and get $P_{m+1}$. Note that there are still very small numerical errors in the projector condition and the energy conservation. To preserve the robustness of our computation, we enforce the projector condition with an error criterium $\delta p=5\times10^{-15}$ for the difference $|P_m^2-P_m|_{ab}<\delta p$. If the criterium is broken, we diagonalize the density matrix $P_{m+1}=U_pDU_p^\dagger$ and correct the diagonal matrix $D\rightarrow(0,0,\dots,0,1,1,\dots,1)$. This step ensures the $N$-state occupation and recovers the projector condition. Note that this enforcement also controls the error in the energy conservation, since it scales similarly to the projector-condition error.
\end{enumerate}

\subsection{Implementation in this work}

\begin{figure}[t]
\centering
\includegraphics[scale=1]{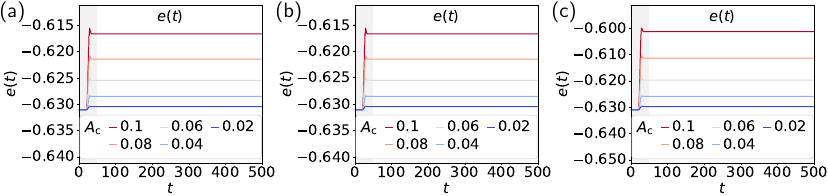}
\caption{\label{suppfig:ee} The energy change in the post-pump dynamics. The time evolution is computed for the linearly polarized pumps with $\mbf e_{\text{p},0}=$ (a) $(1,0,0)$ and (b) $(0,1,0)$, as well as the (c) circularly polarized pump. The center frequency is set as $\omega_\text{c}=2$.}
\end{figure}

We now explain some details of the implementation in our work. Our time evolution starts from the TrH ground state at the $p$-type Van Hove singularity of the kagome lattice. The discrete time step is chosen as $\Delta t=10^{-2}$. At each time step, the errors in the projector condition and the post-pulse-duration energy conservation are at a reasonably small order $O(10^{-12})$. We generally run the time evolution up to $t=500$, meaning $5\times10^4$ iterations for each set of parameters. After the short pulse duration $t\in[0,50]$, the energy change shows a flat plateau under the time-independent Hamiltonian (Fig.~\ref{suppfig:ee}). The energy deviation $e(t=100)-e(t=500)$ in this regime is smaller than the pump-induced energy jump by a scale of $O(10^{-5})$. For example, under the circularly polarized pump with $A_\text{c}=0.1$, where the pump-induced energy jump $\Delta e\approx0.03$ is the largest in our analysis, the energy deviation $e(t=100)-e(t=500)\approx1.2\times10^{-7}$ is comparably small.

\section{Dynamics of charge orders}

The time-dependent Hartree-Fock theory supports an effective analysis of the post-pump dynamics in the kagome metals. To understand the nature of the dynamics, we need to compute some observables that can characterize the state. Since our analysis focuses on the charge orders of the kagome lattice, the observables we need to compute are the charge orders at various momenta. In this section, we discuss the charge orders involved in our analysis and show miscellaneous data that can complement the main text.

\subsection{Types of charge orders}

The charge orders on the kagome lattice can manifest various momenta, forming an enormous group of potential candidates to consider. Nevertheless, in the regimes of our interest, our computation generally finds strong $2\times2$ periodicity in the real-space patterns. This important feature narrows the candidates down to the $2\times1$ ones at $\mbf M$ and the $1\times1$ ones at $\mbf\Gamma$. Since our Hubbard model is nearest-neighbor, we consider only the nearest-neighbor harmonics of the charge orders. These harmonics generally manifest certain symmetry-imposed form factors, and their amplitudes can be determined from the high symmetry points of the Brillouin zone. Therefore, the dynamics of the charge orders can be computed along the time evolution at a reasonable computational expense.

A charge order at a transfer momentum $\mbf q$ can be defined as
\begin{equation}
\Delta_{\mbf k}=\sum_{\tau\tau'}T_{\mbf k\mbf q\tau\tau'}\epvl{c_{(\mbf k+\mbf q)\tau'}^\dagger c_{\mbf k\tau}}.
\end{equation}
Here the sublattice-structure matrix $T_{\mbf k\mbf q}$ is defined from the wavefunctions $\psi_{\mbf k}$ and $\psi_{\mbf k+\mbf q}$ or the chosen site/bond orders. The charge order can be written in terms of an order parameter $ \Delta\in\mathbb C$ and a form factor $f_{\mbf k}\in\mathbb R$ \Scite{lin21prb}
\begin{equation}
\Delta_{\mbf k}=\Delta f_{\mbf k}.
\end{equation}
We want to monitor the time evolution of the order parameter $\Delta$. Since $f_{\mbf k}$ usually takes the form of certain irreducible representation under the $\text{C}_{6v}$ symmetry, we may focus on $N_b$ high-symmetry points $\mbf k_b$ in the Brillouin zone with $f_{\mbf k_b}=\pm1$
\begin{equation}
\Delta
=\frac{1}{N_b}\sum_{b=1}^{N_b}f_{\mbf k_b}\Delta_{\mbf k_b}
=\frac{1}{N_b}\sum_{b=1}^{N_b}f_{\mbf k_b}\sum_{\tau\tau'}T_{b,\tau\tau'}\epvl{c_{(\mbf k_b+\mbf q_b)\tau'}^\dagger c_{\mbf k_b\tau}}.
\end{equation}
Here $\mbf q_b=\pm\mbf q$ are the intra-Brillouin-zone equivalences of the transfer momentum $\mbf q$, and $T_b$ is the sublattice-structure matrix at $\mbf k_b$. A further Fourier transform to the nearest-neighbor pairs $\epvl{i\tau,i'\tau'}$
\begin{equation}
\begin{aligned}
\Delta
&=\frac{1}{N_b}\sum_{b=1}^{N_b}f_{\mbf k_b}\frac{1}{N_\text{BL}}\sum_{\epvl{i\tau,i'\tau'}}T_{b,\tau\tau'}\epvl{c_{i'\tau'}^\dagger c_{i\tau}}e^{i[(\mbf k_b+\mbf q_b)\cdot(\mbf r_{i'\tau'}-\mbf r_0^c)+\mbf k_b\cdot(\mbf r_{i\tau}-\mbf r_0^c)]}\\
&=\frac{1}{N_b}\sum_{b=1}^{N_b}f_{\mbf k_b}\frac{1}{N_\text{BL}}\sum_{\epvl{i\tau,i'\tau'}}T_{b,\tau\tau'}P_{ii'\tau\tau'}e^{i[(\mbf k_b+\mbf q_b)\cdot(\mbf r_{i'\tau'}-\mbf r_0^c)+\mbf k_b\cdot(\mbf r_{i\tau}-\mbf r_0^c)]}
\end{aligned}
\end{equation}
yields the formula for a computation from the density matrix. Note that the center $\mbf r_0^c$ of the zero-th unit cell is taken as the Fourier-transform reference point, which guarantees the $\text{C}_3$ symmetry between the orders. For the charge density and current orders, we further take $\Delta=\text{Re}(\Delta)$ and $\text{Im}(\Delta)$, respectively. Note that the real and imaginary terminologies are defined by their real-space nature. Under a Fourier transform to the momentum space, these two nature may swap in some cases. If this situation occurred, we need multiply the form factor by $i$ to match the real- or imaginary-part extraction.

\begin{figure}[t]
\centering
\includegraphics[scale=1]{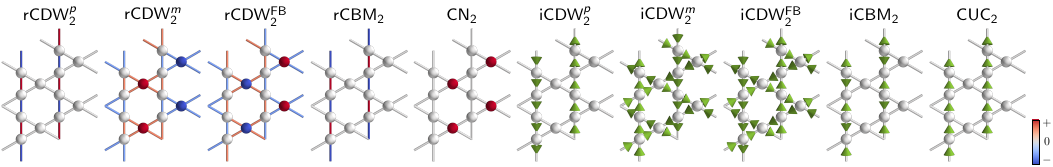}
\caption{\label{suppfig:cos} The real-space patterns of the charge orders in our analysis. The site and bond colors indicate the density deviations from average (see the color bar), while the bond arrows represent the currents.}
\end{figure}

We now consider the charge orders and determine their corresponding setups for the computation. The corresponding real-space patterns are shown for clarity (Fig.~\ref{suppfig:cos}).

\subsubsection{$\mbf M$-point orders}

We first discuss the $\mbf M$-point orders, which correspond to the $2\times1$ CDWs. Here we consider the CDWs in the band basis and focus on the $\mbf M$ points. The wavefunctions at the $p$-type VHS ($p$), the $m$-type VHS ($m$), and the flat band (FB) take the form
\begin{equation}
\psi^p_{\mbf M_a}=[(\delta_{\tau a})_{\tau=0,1,2}]^T,\quad
\psi^m_{\mbf M_a}=[([1-\delta_{\tau a}]/\sqrt2)_{\tau=0,1,2}]^T,\quad
\psi^\text{FB}_{\mbf M_a}=[([1-\delta_{\tau a}][-1]^{(a-\tau)\%3}/\sqrt2)_{\tau=0,1,2}]^T,
\end{equation}
where $\%$ indicates a mod. For a given transfer momentum $\mbf Q_\alpha=\mbf M_{(\alpha+2)\%3}-\mbf M_{(\alpha+1)\%3}\equiv\mbf M_\alpha$, the CDW$_\alpha$ can be either real or imaginary (r or iCDW$^{p,m,\text{FB}}$) \Scite{lin21prb}. Notably, these two orders correspond to distinct form factors $f_{\mbf M_{(\alpha+2)\%3}}=\pm f_{\mbf M_{(\alpha+1)\%3}}=1$ due to $\epvl{c_{\mbf M_{(\alpha+2)\%3}}^\dagger c_{\mbf M_{(\alpha+1)\%3}}}=\epvl{c_{\mbf M_{(\alpha+1)\%3}}^\dagger c_{\mbf M_{(\alpha+2)\%3}}}^*$. Therefore, we may determine the amplitudes of the orders by the $N_b=4$ points $\mbf k_b=\pm\mbf M_{(\alpha+1)\%3}$ and $\pm\mbf M_{(\alpha+2)\%3}$ with $\mbf q_b=\pm\mbf Q_\alpha$ and $\mp\mbf Q_\alpha$. The sublattice-structure matrices are given by the corresponding wavefunctions $T_b=\psi_{\mbf k_b}\psi_{\mbf k_b+\mbf q_b}^T$.

\subsubsection{$\mbf\Gamma$-point orders}

We next discuss the charge orders at $\mbf\Gamma$. These orders are $1\times1$ with zero transfer momentum $\mbf q_b=\mbf q=\mbf0$. Instead of working in the band basis, here we conduct the analysis in the sublattice basis.

The real orders can be either site or bond orders, which correspond to the charge nematicity (CN) and the charge bond modulation (rCBM), respectively. For the site orders, we define three orders CN$_{\alpha=0,1,2}$ by the three sublattices. This definition implies a sublattice-structure matrix $T_b=(\delta_{\tau \alpha}\delta_{\tau'\alpha})_{\tau\tau'}$. Since a uniform site order corresponds to an $s$-wave form factor $f_{\mbf k}=1$, we may take a $N_b=1$ point $\mbf k_b=\mbf\Gamma$. We further subtract the results by the fermion filling $n_\text{f}$, which captures the CN order parameters more appropriately. Meanwhile, rCBM$_{\alpha=0,1,2}$ can be defined by the breathing bond density modulations between the other two sublattices $\tau=(\alpha+1)\%3$ and $\tau=(\alpha+2)\%3$ with the bond vector $\mbf d_\alpha=\mbf r_{0[(\alpha+2)\%3]}-\mbf r_{0[(\alpha+1)\%3]}$. A Fourier transform of the bond order gives
\begin{equation}
\Delta_{\mbf k}\sim(e^{i\mbf k\cdot\mbf d_\alpha}-e^{-i\mbf k\cdot\mbf d_\alpha})=2i\sin(\mbf k\cdot\mbf d_\alpha).
\end{equation}
Since $[\pm\mbf M_{(\alpha+1)\%3}]\cdot\mbf d_\alpha=[\mp\mbf M_{(\alpha+2)\%3}]\cdot\mbf d_\alpha=\pm2\pi/4=\pm\pi/2$, the form factor is maximal at these points with $f_{\mbf M_{(\alpha+1)\%3}}=f_{-\mbf M_{(\alpha+2)\%3}}=-f_{-\mbf M_{(\alpha+1)\%3}}=-f_{\mbf M_{(\alpha+2)\%3}}=1$. We can take these points as the $N_b=4$ momenta $\mbf k_b$. Note that the real order has turned into an imaginary order in the momentum space. To accomodate the imaginary prefactor $i$ in the real-part extraction, we further multiply the form factor by $-i$ as $f_{\mbf M_{(\alpha+1)\%3}}=f_{-\mbf M_{(\alpha+2)\%3}}=-f_{-\mbf M_{(\alpha+1)\%3}}=-f_{\mbf M_{(\alpha+2)\%3}}=-i$. The sublattice-structure matrix is defined as $T_b=[(\delta_{\tau[(\alpha+1)\%3]}\delta_{\tau'[(\alpha+2)\%3]})_{\tau\tau'}-(\delta_{\tau[(\alpha+2)\%3]}\delta_{\tau'[(\alpha+1)\%3]})_{\tau\tau'}]/2$, which antisymmetrizes the complex conjugate pairs under the swap of the sublattices.

The imaginary orders show up as two types of charge bond current orders. The first type is the iCBM, where the staggered currents flow on the bonds along a single direction. The Fourier transform of iCBM$_{\alpha=0,1,2}$
\begin{equation}
\Delta_{\mbf k}\sim i(e^{i\mbf k\cdot\mbf d_\alpha}+e^{-i\mbf k\cdot\mbf d_\alpha})=2i\cos(\mbf k\cdot\mbf d_\alpha)
\end{equation}
is maximal along $\mbf M_\alpha$-$\mbf\Gamma$-$(-\mbf M_\alpha)$ with $\mbf k\cdot\mbf d_\alpha=0$. It is sufficient to take $N_b=1$ point $\mbf k_b=\Gamma$ with $f_{\mbf\Gamma}=1$ and the antisymmetrized sublattice-structure matrix $T_b=[(\delta_{\tau[(\alpha+1)\%3]}\delta_{\tau'[(\alpha+2)\%3]})_{\tau\tau'}-(\delta_{\tau[(\alpha+2)\%3]}\delta_{\tau'[(\alpha+1)\%3]})_{\tau\tau'}]/2$. Another imaginary order is a charge uniform current (CUC), where the uniform current flows on the bonds along a single direction. For CUC$_{\alpha=0,1,2}$, the Fourier transform
\begin{equation}
\Delta_{\mbf k}\sim-i(e^{i\mbf k\cdot\mbf d_\alpha}-e^{-i\mbf k\cdot\mbf d_\alpha})=2\sin(\mbf k\cdot\mbf d_\alpha)
\end{equation}
gives the real version of rCBM$_\alpha$. We take the same setup as the one for rCBM$_\alpha$, except for the opposite definition of the form factor $f_{\mbf M_{(\alpha+1)\%3}}=f_{-\mbf M_{(\alpha+2)\%3}}=-f_{-\mbf M_{(\alpha+1)\%3}}=-f_{\mbf M_{(\alpha+2)\%3}}=i$ and the symmetrized version of the sublattice-structure matrix $T_b=[(\delta_{\tau[(\alpha+1)\%3]}\delta_{\tau'[(\alpha+2)\%3]})_{\tau\tau'}+(\delta_{\tau[(\alpha+2)\%3]}\delta_{\tau'[(\alpha+1)\%3]})_{\tau\tau'}]/2$.

\subsection{Computation of frequency spectrum}

\begin{figure*}[t]
\centering
\includegraphics[scale=1]{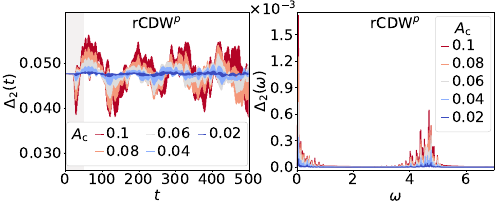}
\caption{\label{suppfig:freq} The raw time evolution and the frequency spectrum of rCDW$^p$ under the pumps with $\omega_\text{c}=2$ and various $A_\text{c}$'s.}
\end{figure*}

In the post-pump dynamics, the charge orders experience strong oscillations under the time evolution. To analyze these oscillations, we conduct a Fourier transform to obtain the frequency spectrum. Focusing on the post-pump dynamics, we only consider the order parameters $\Delta_\alpha(t)$ of each charge order after the pulse duration $t>50$. The data are refined $\tilde\Delta_\alpha(t)=\Delta_\alpha(t)-f(t)$ by subtracting a linear regression fitting $f(t)=at+b$, which removes the gradual decay after the pump. A Fourier transform to the frequency space
\begin{equation}
\Delta_\alpha(\omega)=\frac{1}{450}\int_{50}^{500}dt\tilde\Delta_\alpha(t)e^{i\omega(t-50)}
\end{equation}
then produces the frequency spectrum of the charge-order dynamics.

As an example, we take the rCDW$^p$ under the linearly polarized pump with $\omega_\text{c}=2$ (Fig.~\ref{suppfig:freq}). From the raw time evolution, we can see a combination of high- and low-frequency oscillations. The corresponding frequency spectrum shows two bundles. The high-frequency bundle at $4\lesssim\omega\lesssim6$ corresponds to the rapid breaking and recombination of the particle-hole condensates. Meanwhile, the low-frequency bundle at $0\leq\omega\lesssim1$ corresponds to the coherent oscillations of the charge-order collective modes. Note that this two-bundle feature is generic for the relevant charge orders we consider in the main text, including the r and iCDWs, and CN. The iCBM also manifests the two-bundle oscillations, although at much weaker strengths. On the other hand, the rCBM and the CUC show a single bundle at $2\lesssim\omega\lesssim3$ of rapid oscillations around zero.

\subsection{Miscellaneous dynamics}

In the main text, we have presented the primary results of the post-pump dynamics. In this section, we present more results which can complement the figures in the main text.

\subsubsection{Linearly polarized pump}

\begin{figure*}[t]
\centering
\includegraphics[scale=1]{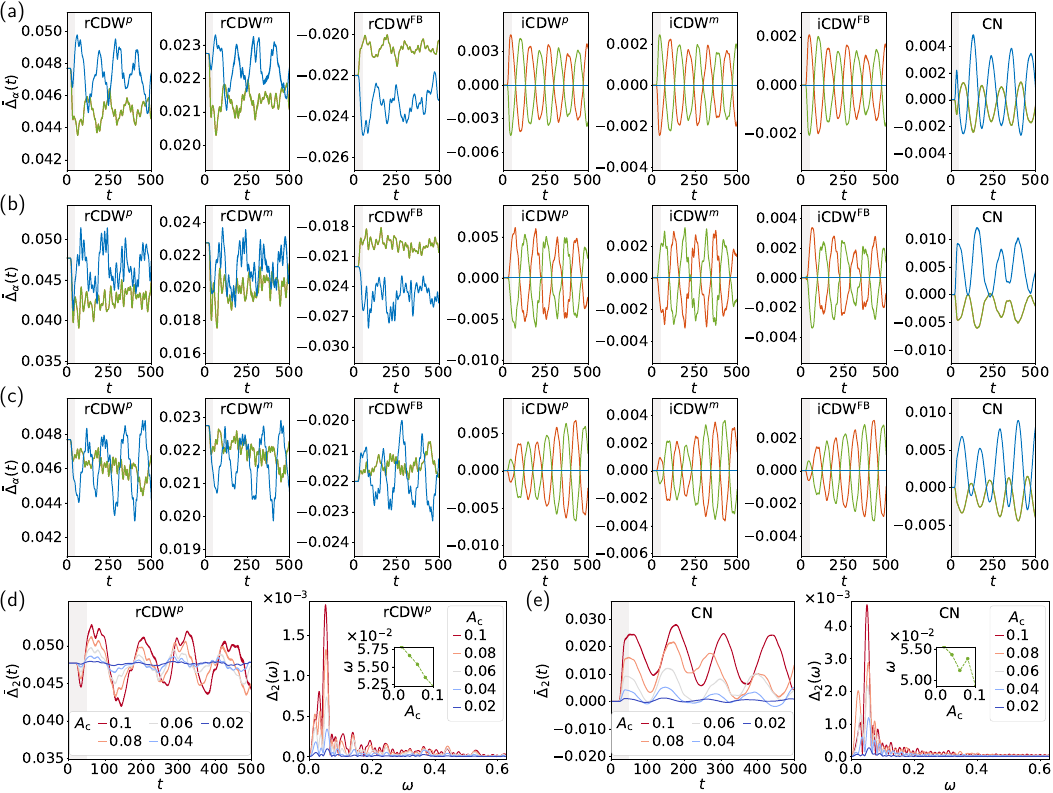}
\caption{\label{suppfig:lin100} Post-pump dynamics under the linearly polarized pumps with $\mbf e_{\text{p},0}=(1,0,0)$. The dynamics under the pumps with (a) $\omega_\text{c}=2$, (b) $\omega_\text{c}=2.5$, and (c) $\omega_\text{c}=3$ are presented, where $A_\text{c}=0.06$ is chosen in all cases. The $\alpha=0,1$ orders evolve identically in rCDW$^{p,m,\text{FB}}$ and CN, while the $\alpha=2$ order undergoes a different evolution. The $A_\text{c}$ dependence is analyzed for (d) the rCDW$^p$ with $\omega_\text{c}=2$ and (e) the CN with $\omega_\text{c}=2.5$, where the insets show the trend of the major-peak frequency with increasing $A_\text{c}$. The relevant time and frequency scales under our unit conversion are pulse duration (gray time domain) $t=50\equiv32.9\text{ fs}$, total time $t=500\equiv0.329\text{ ps}$, and major-peak frequency scale $\omega=0.05\equiv12.085\text{ THz}$.}
\end{figure*}

\begin{figure*}[t]
\centering
\includegraphics[scale=1]{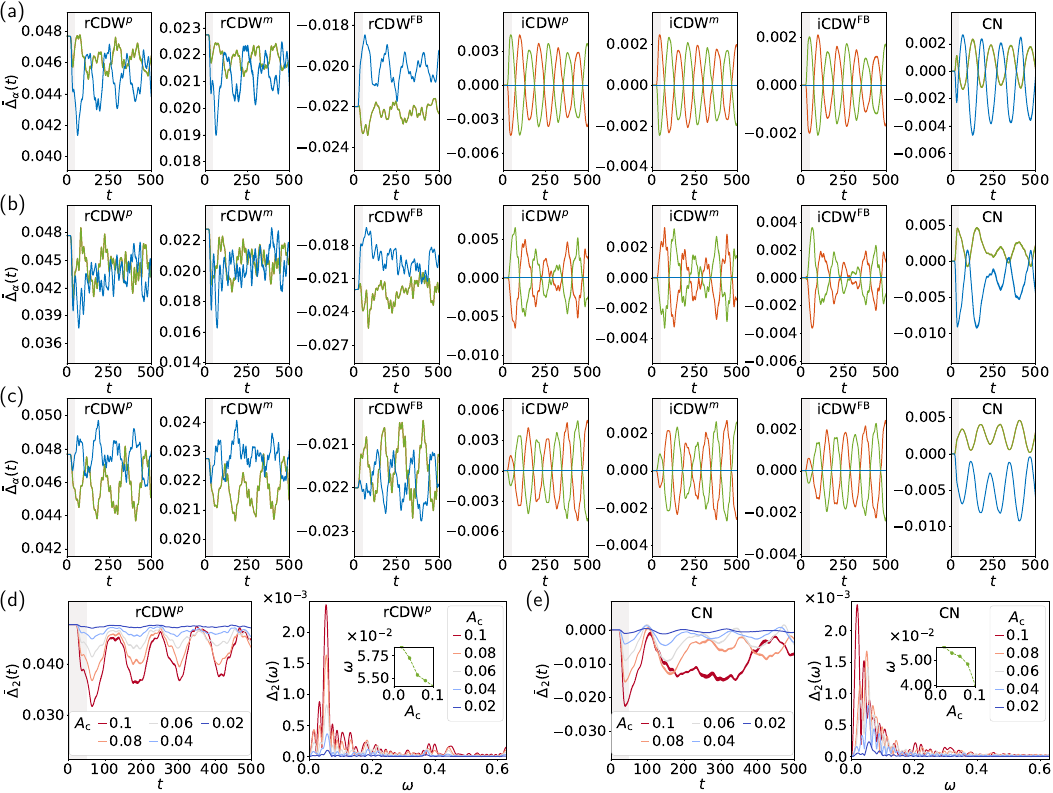}
\caption{\label{suppfig:lin010} Post-pump dynamics under the linearly polarized pumps with $\mbf e_{\text{p},0}=(0,1,0)$. The dynamics under the pumps with (a) $\omega_\text{c}=2$, (b) $\omega_\text{c}=2.5$, and (c) $\omega_\text{c}=3$ are presented, where $A_\text{c}=0.06$ is chosen in all cases. The $\alpha=0,1$ orders evolve identically in rCDW$^{p,m,\text{FB}}$ and CN, while the $\alpha=2$ order undergoes a different evolution. The $A_\text{c}$ dependence is analyzed for (d) the rCDW$^p$ with $\omega_\text{c}=2$ and (e) the CN with $\omega_\text{c}=2.5$, where the insets show the trend of the major-peak frequency with increasing $A_\text{c}$. The relevant time and frequency scales under our unit conversion are pulse duration (gray time domain) $t=50\equiv32.9\text{ fs}$, total time $t=500\equiv0.329\text{ ps}$, and major-peak frequency scale $\omega=0.05\equiv12.085\text{ THz}$.}
\end{figure*}

We first present the results for the linearly polarized pumps with $\mbf e_{\text{p},0}=(1,0,0)$ (Fig.~\ref{suppfig:lin100}) and $(0,1,0)$ (Fig.~\ref{suppfig:lin010}). As discussed in the main text, there exist the directional preferences of the rCDWs, the enhancement on the flat band, and the resonant enhancement of the CN at $\omega_\text{c}\approx2.5$. Notably, over the resonant frequency $\omega_\text{c}>2.5$, the directional preferences of the rCDWs seem to become opposite in the initial period after the pump. This opposite behavior changes back to the original one relatively fast under the time evolution. On the other hand, the iCDW$_{0,1}$s show strong oscillations with zero time averages, while the iCDW$_2$s remain zero under the time evolution. This behavior also shows up in the iCBM at much weaker strengths.

We also analyze the dependence of the post-pump dynamics on the center amplitude $A_\text{c}$. In particular, our interest lies in how the major peak shifts with increasing $A_\text{c}$. Note that the strong pumps may generate stronger irregular features than the regular oscillations observed under the weak pumps. To focus on these regular oscillations, we extract the peak frequencies in the regime $[(1/2)\omega^{0.02}_\text{peak},(3/2)\omega^{0.02}_\text{peak}]$ around the major peak $\omega=\omega^{0.02}_\text{peak}$ with $A_\text{c}=0.02$. Our analysis finds a general softening of the charge-orde collective modes with increasing $A_\text{c}$.

\subsubsection{Circularly polarized pump}

\begin{figure*}[t]
\centering
\includegraphics[scale=1]{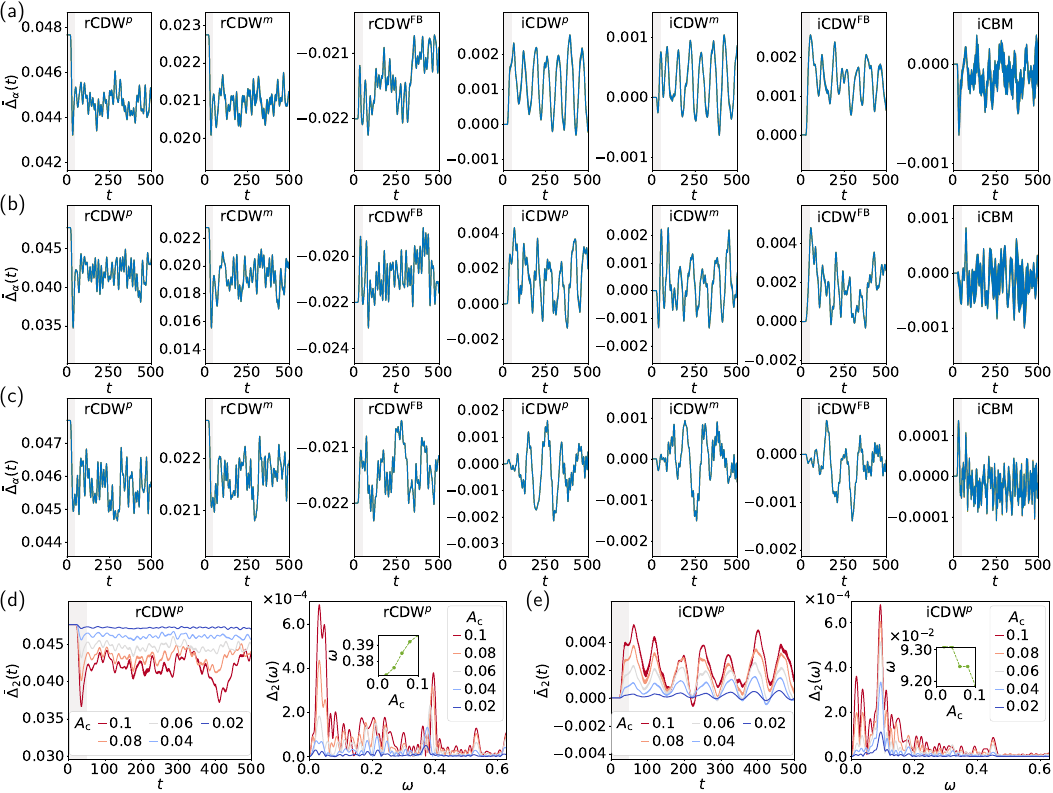}
\caption{\label{suppfig:cir} Post-pump dynamics under the circularly polarized pumps. The dynamics under the pumps with (a) $\omega_\text{c}=2$, (b) $\omega_\text{c}=2.5$, and (c) $\omega_\text{c}=3$ are presented, where $A_\text{c}=0.06$ is chosen in all cases. In each case, the smeared time evolution of the $\alpha=0,1,2$ orders are nearly identical. The $A_\text{c}$ dependence is analyzed for (d) the rCDW$^p$ and (e) the iCDW$^p$ with $\omega_\text{c}=2$, where the insets show the trend of the major-peak frequency with increasing $A_\text{c}$. The relevant time and frequency scales under our unit conversion are pulse duration (gray time domain) $t=50\equiv32.9\text{ fs}$, total time $t=500\equiv0.329\text{ ps}$, and major-peak frequency scales $\omega=0.39\equiv94.263\text{ THz}$ and $\omega=0.09\equiv21.753\text{ THz}$.}
\end{figure*}

We next present the results for the circularly polarized pumps (Fig.~\ref{suppfig:cir}). As discussed in the main text, there exist the uniform suppressions of the rCDWs and the emergence of the iCDWs. Notably, over the resonant frequency $\omega_\text{c}>2.5$, the iCDWs seem to gain opposite time averages. On the other hand, the pump pulse also triggers a much weaker iCBM, whose loop current order also breaks the time-reversal symmetry.

We also analyze the dependence of the post-pump dynamics on the center amplitude $A_\text{c}$. Interestingly, the rCDW$^p$ shows an opposite behavior to the softening, where the regular-oscillation peak shifts toward higher frequency with increasing $A_\text{c}$. On the other hand, the iCDW$^p$ shows a softening with increasing $A_\text{c}$.

\end{document}